\documentstyle[preprint,aps,tighten,epsf]{revtex}
\begin{document}
\draft
\def\etal{{\em et\ al.\/} }
\def\eg{{\em e.g.\/}}
\def\ie{{\em i.e.\/}}
\def\beq{\begin{equation}}
\def\eeq{\end{equation}}
\def\beqarr{\begin{eqnarray}}
\def\eeqarr{\end{eqnarray}}
\def\ix{\int_{-L/2}^{L/2}dx\,}
\def\ixi{\int dx\,}
\def\Zint{
  {\mathchoice
    {{\sf Z\hskip-0.45em{}Z}}
    {{\sf Z\hskip-0.44em{}Z}}
    {{\sf Z\hskip-0.34em{}Z}}
    {{\sf Z\hskip-0.35em{}Z}}
    }
  }
\def\delx{\partial_x}
\def\delxp{\partial_{x'}}
\def\delt{\partial_{\tau}}
\def\delT{\partial_t}
\def\XLL{{$\chi$LL}}
\def\gtlt{\,{\stackrel {\scriptscriptstyle >}{\scriptscriptstyle <}}\,}
\def\bh{{\hat \beta}}
\def\ph{{\hat \phi}}
\def\XSG{{$\chi$SG }}
\def\bJ{{\bf J}}
\def\Lt{L^{(1/2)}}
\def\Lh{L'^{(1/2)}}
\def\lh{{\hat \lambda}}
\def\nb{{\bar \nu}}
\def\tb{{\bar \tau}}
\def\ID{I}
\def\Ab#1{{\bf A}^{(#1)}}
\def\Bb#1{{\bf B}^{(#1)}}
\def\Cb#1{{\bf C}^{(#1)}}
\def\Af#1{\tilde{\bf A}^{(#1)}}
\def\Bf{\tilde{\bf B}}
\def\Cf#1{\tilde{\bf C}^{(#1)}}
\def\Rf#1{\tilde{\bf R}^{(#1)}}
\def\Gf#1{\tilde{\bf G}^{(#1)}}
\def\X#1{2\pi(v_{#1}\tau-ix)}
\preprint{IASSNS-HEP-99/76}
\title{
  Quantum Hall Bilayers and the Chiral Sine-Gordon Equation}

\author{J. D. Naud$^1$\footnote{Corresponding Author: naud@princeton.edu, 609.258.5983 (Tel),
609.258.1006 (FAX)}, Leonid P. Pryadko$^2$ and S. L. Sondhi$^1$}

\address{
$^1$Department of Physics,
Princeton University,
Princeton, NJ 08544, USA
}

\address{
$^2$School of Natural Sciences,
Institute for Advanced Study,
Olden Lane,
Princeton, NJ 08540}
\date{December 2, 1999}
\maketitle

\begin{abstract}

  The edge state theory of a class of symmetric double-layer quantum
  Hall systems with interlayer electron tunneling reduces to the sum
  of a free field theory and a field theory of a chiral Bose field
  with a self-interaction of the sine-Gordon form.  We argue that the
  perturbative renormalization group flow of this chiral sine-Gordon
  theory is distinct from the standard (non-chiral) sine-Gordon
  theory, contrary to a previous assertion by Renn, and that the
  theory is manifestly sensible only at a discrete set of values of
  the inverse period of the cosine interaction ($\bh$). We obtain exact
  solutions for the spectra and correlation functions of the chiral
  sine-Gordon theory at the two values of $\bh$ at which electron
  tunneling in bilayers is {\em not\/} irrelevant. Of these, the marginal
  case ($\bh^2=4$) is of greatest interest: the spectrum of the
  interacting theory is that of two Majorana fermions with different,
  dynamically generated, velocities. For the experimentally observed
  bilayer 331 state at filling factor $1/2$, this implies the {\em
  trifurcation\/} of electrons added to the edge.  We also present a
  method for fermionizing the theory at the discrete points
  ($\bh^2\in\Zint^{+}$) by the introduction of auxiliary degrees of
  freedom that could prove useful in other problems involving quantum
  Hall multi-layers.

\end{abstract}

\section{Introduction}
\label{sec:intro}

This paper is concerned with a chiral version of the celebrated
sine-Gordon (SG) field theory in one dimension, in which the standard
kinetic term is replaced by the kinetic term for a {\em chiral\/} Bose 
field. More precisely, we are concerned with the Hamiltonian for a
periodic boson of radius $R=1/\bh$,
\begin{equation} 
  \label{eq:general-chiSG-hamiltonian}
  {\cal H}_{\chi
    SG}[\phi]=\int dx\, \left[{{1\over 4\pi}(\delx \phi)^2 - \kappa
    \cos\left({\bh \phi}\right)}\right],  
\end{equation} 
supplemented by the equal-time commutation relation, 
\begin{equation}
  [\phi(x),\phi(x')]=i\pi \,{\rm sgn}(x-x'), 
\end{equation} 
appropriate for a chiral Bose field. The Hamiltonian is suggestive of
the standard sine-Gordon theory, but the resemblance is misleading and
the physics, as we shall see, is very different on account of the
chiral nature of the field. Indeed, a previous analysis of this
problem by Renn \cite{renn} is in error precisely because of his
neglect of the chiral constraint. We will comment further on the
difference between the two theories at several points in the main
text.  

The chief interest of the chiral sine-Gordon ($\chi$SG) theory is that
it arises naturally, as we detail below, in the edge theory of
double-layer quantum Hall systems when it is necessary to take account
of weak interlayer-tunneling within Wen's chiral bosonic description
of the edge dynamics. As such it is the simplest of a family of
interacting chiral theories in one dimension that describe the
dynamics of multicomponent systems.  (Such systems yield non-chiral
problems as well, in cases where there are counter-propagating modes,
and they have attracted considerable attention
\cite{kfp,haldane,moore&wen}.) These theories are much more
constrained than generic non-chiral theories, but they are by no means
trivial and they exhibit unusual connections between renormalization
group flows and spectral rearrangements. In a sense, they are
intermediate in complexity between quantum impurity problems, which
can be mapped to chiral problems with interactions localized at a
point, and generic one dimensional field theories, and seem worth
studying on this account as well.  Dijkgraaf has explored some general
properties of chiral theories from a conformal field theory
viewpoint\cite{Dijkgraaf}, and several authors have considered
particular examples such as the chiral Potts model
\cite{Cardy} and the chiral Schwinger model\cite{raj&jackiw}.
Finally, the study of chiral field theories is not without interest in
another condensed matter context: Ho and Coleman have recently
motivated the study of interacting chiral Majorana theories by
appealing to the phenomenology of the cuprates and have presented
solutions of several models by an interesting ``fermionic bootstrap''
\cite{Ho&Coleman}.

We begin with a review of the bosonic formulation of the edge theory
of double-layer quantum Hall systems with interlayer tunneling
(Section \ref{sec:edgetheory}).  For systems with one mode per layer
we show that for clean, symmetric realizations, the edge theory
separates into the sum of a free boson Hamiltonian and the \XSG
Hamiltonian with $\bh^2\in\Zint^{+}_{\rm even}$ (the ``bosonic''
sequence).  We argue that the \XSG theory is manifestly sensible only
for integer values of $\bh^2$.  Layer asymmetry and disorder, which
lead to a more complicated dynamics, will be discussed elsewhere
\cite{nps-wip}.  Next we consider the renormalization group flow of
the \XSG theory, perturbatively in $\lambda$, and argue that it
differs from the usual SG theory (Section \ref{sec:RG}).  We argue
that $\bh$ is not renormalized and for $\bh^2=4$ there is a line of
fixed points parameterized by $\lambda$.  In Section
\ref{sec:exactsolns} we obtain exact solutions for the spectrum and
correlation functions of the \XSG theory for $\bh^2=2$ and $\bh^2=4$
which are the two members of the bosonic sequence at which the
interaction is not irrelevant. At $\bh^2=2$ we show by
``re-bosonization'', that the spectrum of the interacting theory is,
in the thermodynamic limit, still that of a free chiral boson.
Perhaps of greatest interest is our solution by fermionization at
$\bh^2=4$, relevant to the experimentally observed 331 state, where
the \XSG theory is shown to be equivalent to a theory with two chiral
Majorana fermions with dynamically generated distinct velocities.
Next we present a method for fermionizing the
\XSG theory at $\bh^2\in\Zint^{+}$ by the introduction of auxiliary bosons
(Section \ref{sec:fermionizations}).  We show that for $\bh^2=2$ and
$\bh^2=4$ the fermionized theory with auxiliary degrees of freedom can
be solved exactly, and after projection onto the \XSG Hilbert space we
recover the results of Section \ref{sec:exactsolns}.  In Section
\ref{sec:fermiseq} we briefly consider the \XSG theory at the points
$\bh^2\in\Zint^{+}_{\rm odd}$ (the ``fermionic'' sequence), which do not
correspond to any double-layer system.  We comment briefly on the
difficulties in understanding our results from the kinds of
semiclassical considerations that are so useful in the standard
sine-Gordon problem in Section \ref{sec:semiclass}. We conclude with a
discussion (Section \ref{sec:disc}) and several Appendices giving
detailed calculations in support of statements made in the text.

\section{Edge Theory of Double-layer Systems}
\label{sec:edgetheory}
\subsection{Double-layer Systems without Tunneling}

\begin{figure}[htbp]
  \begin{center}
    \leavevmode
    \epsfxsize=0.4\columnwidth
    \epsfbox{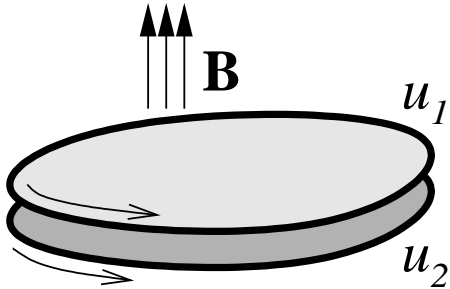}\vskip1mm
    \caption{The overall geometry of a double-layer quantum Hall system, with
      edge modes in both layers ($u_i$) propagating in the same direction.
      The chiral sine-Gordon theory describes the dynamics of the
      neutral mode in the presence of uniform interedge electron tunneling.}
    \label{fig:qh-cartoon}
  \end{center}
\end{figure}

We are interested in a system consisting of two parallel layers of
2DEGs (two-dimensional electron gases) in a strong perpendicular
magnetic field, with a confining potential that restricts the
electrons in each layer to a region with the topology of a disc.
The geometry is sketched in Fig.~\ref{fig:qh-cartoon}.
The simplest Abelian quantum Hall states in double-layer systems are well
described by the two-component generalization of the Laughlin
wavefunction first introduced by Halperin\cite{halperin}:
\begin{equation}
  \label{eq:wavefnc}
  \Psi_{m,m',n}(\{z_{i
    \alpha}\})=\prod_{\alpha<\beta}(z_{1\alpha}-z_{1\beta})^m
  (z_{2\alpha}-z_{2\beta})^{m'} \prod_{\alpha,\beta}
  (z_{1\alpha}-z_{2\beta})^n\,\, e^{-\sum_{i,\alpha}|z_{i \alpha}|^2/4},
\end{equation}
where $z_{i \alpha}$ is the complex coordinate of electron $\alpha$ in
layer $i$, and we work in the symmetric gauge.  Here $m$ and $m'$ are
positive odd integers characterizing the intralayer correlations and
$n$ is a non-negative integer specifying interlayer correlations.  The
filling factor in each layer is\cite{gm1} 
\begin{equation}
  \label{eq:fillings}
  \pmatrix{\nu_1 \cr \nu_2 \cr}={1\over (mm'-n^2)}\pmatrix{m'-n \cr
    m-n \cr}. 
\end{equation}
Following convention we refer to these states by the shorthand
$mm'n$.

The gapless excitations of the system are confined to the edges of the
droplets, and for unreconstructed edges they are described at
low-energies by the effective Hamiltonian\cite{wen}
\begin{equation}
  \label{eq:H_0}
  {\cal H}_0=\ix {1\over{4\pi}}V_{ij} 
  \,:\!\delx u_i\,\delx u_j\!:,
\end{equation}
where we have assumed the long-range part of the Coulomb interaction
is screened.  Here $x$ is the coordinate along the edge of length $L$
and normal ordering will be defined below.  $V$ is a symmetric matrix
whose elements depend on the details of the confining potentials and
the interactions at the edge.  We require $V$ to be positive definite
so that the Hamiltonian is bounded from below.  We can parameterize
this matrix by
\begin{equation}
  \label{eq:V}
  V=\pmatrix{ v & g \cr g & v' \cr},\quad \hbox{where}\quad g^2 < vv'.
\end{equation}
The $u_i(t,x)$ are chiral Bose fields with compactification
radius $R=1$ (\ie, $ u \sim  u + 2\pi$), which satisfy the
equal-time commutation relations:
\begin{equation}
  \label{eq:varphi_CR}
  [ u_i(t,x), u_j(t,x')]=i\pi K_{ij}\,{\rm sgn}(x-x'),
\end{equation}
where $K$ is a symmetric, integer-valued matrix which characterizes
the topological properties of the system and is completely determined by
the exponents in the bulk wavefunction\cite{wen&zee}
\begin{equation}
  \label{eq:K}
  K=\pmatrix{ m & n \cr n & m' \cr}.
\end{equation}
From here on, any expression in which the time arguments of the
operators are suppressed is to be understood to hold at equal times.
Eq.~(\ref{eq:varphi_CR}) is valid in the limit of large $L$; if we
retained the full $L$ dependence the r.h.s.\ would be periodic 
(${\rm mod}\; 2\pi iK_{ij}$) in $(x-x')$ with period $L$.

Since the chiral Bose fields are compact, we can introduce
the integer-valued Hermitian topological charge operators
\begin{equation}
  \label{eq:topcharge}
  {\cal N}_i={1\over 2\pi}\ix \delx u_i(x),
\end{equation}
which by Eq.~(\ref{eq:varphi_CR}) obey:
\begin{equation}
  \label{eq:N,varphi_CR}
  [{\cal N}_i, u_j(x)]=iK_{ij},\qquad [{\cal N}_i,{\cal N}_j]=0.
\end{equation}
For a chiral boson with a radius $R\neq 1$, the coefficient of the
integral in Eq.~(\ref{eq:topcharge}) would be $1/(2\pi R)$.  The
electric charge density on the edge of layer $i$, $\rho_i(x)$, and the
total charge, $Q_i$, are related to the corresponding topological
quantities by:
\begin{equation}
  \label{eq:rho,Q}
  \rho_i(x)={1\over 2\pi}K^{-1}_{ij}\delx u_j(x),\qquad
  Q_i=K^{-1}_{ij}{\cal N}_j. 
\end{equation}
The vertex operators which create an edge quasiparticle or an edge
electron in layer $i$ are, respectively,
\begin{equation}
  \label{eq:vertex}
  \Psi_{qp,i}^{\dagger}(x)\propto e^{-iK^{-1}_{ij} u_j(x)},
  \qquad\qquad 
  \Psi_{i}^{\dagger}(x)\propto e^{-i u_i(x)}.
\end{equation}
These operators change the topological and electrical charges
according to
\beqarr
\left[{{\cal N}_i,\Psi_{qp,j}^{\dagger}}\right]&=&
\delta_{ij}\Psi_{qp,j}^{\dagger},\qquad 
\left[{Q_i,\Psi_{qp,j}^{\dagger}}\right]=K^{-1}_{ij}
\Psi_{qp,j}^{\dagger}, \nonumber \\
\left[{{\cal N}_i,\Psi_j^{\dagger}}\right]&=&K_{ij}
\Psi_j^{\dagger},\qquad 
\left[{Q_i,\Psi_j^{\dagger}}\right]=\delta_{ij}
\Psi_j^{\dagger}. \label{eq:charges}
\eeqarr
The physical meaning of the edge quasiparticle creation operator
is as follows.  When a quasihole is created in the bulk fluid (\eg, by
the insertion of a flux quantum), the charge depleted from the
vicinity of the quasihole accumulates at the boundary of the droplet.
In the somewhat artificial theoretical construction in which the ``bulk''
and ``edge'' of the droplet are considered distinct, this accumulation
appears as the addition of a fractional amount of charge to the edge
theory, which is accomplished by the edge quasiparticle creation
operator.

\begin{figure}[htbp]
  \begin{center}
    \leavevmode
    \epsfxsize=0.5\columnwidth
    \epsfbox{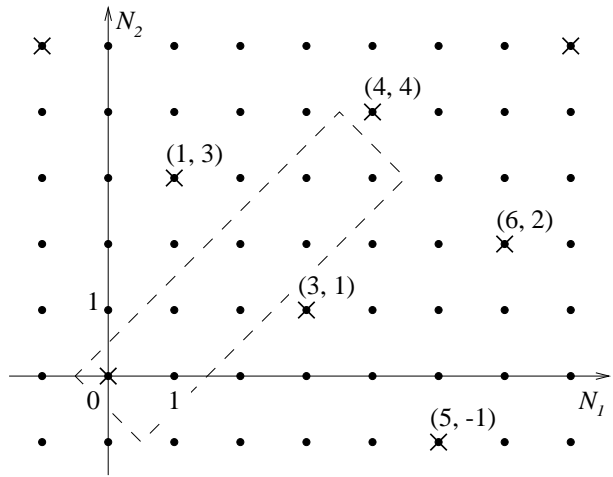}    \vskip2mm
    \caption{The topological charge lattice in the 
    ${\cal N}_1{\cal N}_2$-plane for the double-layer 331 state.  
    The crosses denote points in the electron sublattice (\ie, sector)
    containing the origin.  The unit cell of this sublattice is 
    enclosed by the dashed line.}
    \label{fig:lattice}
  \end{center}
\end{figure}

If we consider the action of the vertex operators (\ref{eq:vertex}) in
a two-dimensional integer lattice in the ${\cal N}_1{\cal N}_2$-plane,
the quasiparticle creation operators can be identified with the basis
vectors of this lattice: $\Psi_{qp,1}^{\dagger}\rightarrow (1,0),
\Psi_{qp,2}^{\dagger}\rightarrow (0,1)$.  The vectors corresponding to
the electron creation operators, $\Psi_1^{\dagger}\rightarrow (m,n),
\Psi_2^{\dagger}\rightarrow (n,m')$, are the basis vectors of an
``electron sublattice'' consisting of all states connected to one
another by adding or removing only electrons from the
edges\cite{milo&read}.  The number of distinct electron sublattices is
equal to the volume of the unit cell spanned by the electron vectors:
$mm'-n^2$.  We shall refer to these as different sectors of the edge
theory.  Note that the discrete spectrum of the topological charges also places 
restrictions on the allowed vertex operators, as we describe below.
An example of the topological charge lattice is shown in 
Fig.~\ref{fig:lattice} for the 331 state (\ie, $m=m'=3,n=1$).  

From Eqns.~(\ref{eq:N,varphi_CR}) and (\ref{eq:vertex}) we find
\beqarr
\Psi_i^{\dagger}(x+L)&=&\Psi_i^{\dagger}(x)
e^{-2\pi i {\cal N}_i}e^{-i\pi K_{ii}}, \nonumber \\
\Psi_{qp,i}^{\dagger}(x+L)&=&\Psi_{qp,i}^{\dagger}(x)e^{-2\pi i
  K^{-1}_{ij}{\cal N}_j} 
e^{-i\pi K^{-1}_{ii}}. \label{eq:BC} 
\eeqarr
Since ${\cal N}_i \in\Zint$ the boundary conditions on the electron
operators are independent of the values of the topological charges.
Within the sector containing the origin, ${\cal N}_i=0$, the same
statement holds for the quasiparticle operators.  However, in other
sectors (\eg, the state with a single quasiparticle in layer $1$,
${\cal N}_1=1,{\cal N}_2=0$) the quasiparticle operator acquires an
additional phase as it is taken around the edge.  Since creating a
quasiparticle at the edge requires the creation of a quasihole in the
bulk as noted above, this additional phase can be
interpreted as the Aharonov-Bohm phase due to the flux creating the
bulk quasihole\cite{Sandler}.

We are interested in the case where the two edge modes have the same
chirality.  Since $V$ is positive definite, the chirality of the edge
bosons is determined by the signs of the eigenvalues of $K$.
Therefore we shall restrict our discussion to the ``maximally chiral''
case in which both eigenvalues of $K$ are positive, which requires
$n^2<mm'$.  (The requirement that the filling factor in each layer be
positive then gives $n<m,m'$.)  In this case, the scaling dimensions
of vertex operators are independent of $V$\cite{kane&fisher}.
Furthermore we shall consider {\em only\/} the symmetric case: $m=m'$
and $v=v'$.  Our result in Section \ref{sec:RG} for the lowest-order
perturbative RG flow of the tunneling amplitude applies for the case
$v\neq v'$ and can easily be extended to states with $m\neq m'$, but
the mapping to the chiral sine-Gordon theory exists only in the
symmetric case, as we demonstrate below.

The Hamiltonian, Eq.~(\ref{eq:H_0}), can be simplified by performing
an orthogonal transformation on the chiral Bose fields to diagonalize
$K$ and $V$, followed by a rescaling to bring $K$ to the identity.
The combined transformation to the new fields $\phi_{\rm c},\phi_{\rm n}$ is:
\begin{equation}
  \label{eq:rotatebosons}
   \pmatrix{u_1 \cr u_2 \cr}={1\over\sqrt{2}}\pmatrix{\sqrt{m+n} &
    -\sqrt{m-n} \cr \sqrt{m+n} & \sqrt{m-n} \cr}
   \pmatrix{\phi_{\rm c} \cr \phi_{\rm n} \cr}.
\end{equation}
We will refer to $\phi_{\rm c},\phi_{\rm n}$ as the charged and neutral edge
modes, respectively, since the topological charge of $\phi_{\rm c}$ is
proportional to the total electric charge on the two edges while the
topological charge of $\phi_{\rm n}$ is proportional to the electric charge
difference between the edges.  When we write $\phi_{\rm i}$, the index
${\rm i}$ 
runs over the two values ${\rm i=c,n}$.  In terms of the new fields, the
commutators are now independent and conventionally normalized:
\begin{equation}
  \label{eq:phi_CR}
  [\phi_{\rm i}(x),\phi_{\rm j}(x')]=i\pi \delta_{\rm ij}\,{\rm sgn}(x-x').
\end{equation}
The mode expansions of the rotated chiral bosons are
\begin{equation}
  \label{eq:phi_modes}
  \phi_{\rm i}(x)={2 \pi \over L}N_{\rm i} R_{\rm i} x -{1\over R_{\rm
  i}}\chi_{\rm i} + 
  \sum_{r=1}^{\infty}{1\over\sqrt{r}} 
  e^{-k_r a/2}(e^{-ik_r x}b_{{\rm i}r}^\dagger+e^{ik_r x}b_{{\rm i}r}),
\end{equation}
where $k_r\equiv 2\pi r/L$, $R_{\rm i}$ are the compactification radii of
the rotated fields, $N_{\rm i}$ are the new topological charges, $\chi_{\rm i}$
are the conjugate Hermitian phase operators, $b_{{\rm i}r}$ are bosonic
annihilation operators, and $a$ is a short-distance cutoff.  The only
nonvanishing commutators are
\begin{equation}
  \label{eq:CR_modes}
  [b_{{\rm i}r},b_{{\rm j}s}^{\dagger}]=\delta_{\rm ij}\delta_{rs},\qquad
  [\chi_{\rm i},N_{\rm j}]=i\delta_{\rm ij}. 
\end{equation}
The normal ordering introduced in Eq.~(\ref{eq:H_0}) is defined with
respect to the bosonic oscillators $b_{{\rm i}r},b_{{\rm i}r}^{\dagger}$.
Up to this point we have neglected normal ordering when writing vertex
operators.  Using Eq.~(\ref{eq:phi_modes}) one can show that normal
ordering produces a multiplicative renormalization of vertex operators
which to lowest order in $a/L$ gives
\begin{equation}
  \label{eq:normal-ordering}
  :\!e^{i \alpha \phi_{\rm i}(x)}\!:\,= \left({L\over 2\pi
      a}\right)^{\Delta(\alpha)}e^{i\alpha\phi_{\rm i}(x)},
\end{equation}
where $\Delta(\alpha)\equiv \alpha^2/2$ is the scaling dimension of
the operator, and also its conformal spin since $\phi_{\rm i}$ is
a chiral field.

Using Eqns.~(\ref{eq:rotatebosons}) and (\ref{eq:phi_modes}) we can
express the Hamiltonian (\ref{eq:H_0}) in terms of the modes of the
uncoupled fields:
\begin{equation}
  \label{eq:H_0modes}
  {\cal H}_0=\ix \left[{{1\over{4\pi}}v_{\rm i}
        \,:\!\!(\delx\phi_{\rm i})^2\!:\,
        }\right]
  ={\pi v_{\rm i}\over L}(R_{\rm i} N_{\rm i})^2 + {2\pi\over
        L}\sum_{r=1}^{\infty}r v_{\rm i} 
  b^{\dagger}_{{\rm i}r}b_{{\rm i}r}, 
\end{equation}
where the velocities of the eigenmodes are $v_{\rm c,n}=(m\pm n)(v\pm
g)$.  The structure of the Hilbert space is now clear.  There are an
infinite number of states $|N_{\rm c},N_{\rm n}\rangle$ which are
eigenstates of the topological charge operators and which are
annihilated by $b_{{\rm i}r}$ for ${\rm i=c,n}$, and $r>0$.  This
infinite set of oscillator vacua is divided into a finite number of
sectors as defined above.  The (zero-energy) ground state is
$|0,0\rangle$.  Note that because Eq.~(\ref{eq:rotatebosons}) defines
a homogeneous linear transformation, the states $|N_{\rm c},N_{\rm
  n}\rangle$ are also vacua for the oscillator modes of the original
bosons ($u_i$).

The values of the radii $R_{\rm i}$ are determined by the requirement
that the new winding number operators $N_{\rm i}$ have integer
eigenvalues.  From the zero-mode piece of the inverse of the
transformation (\ref{eq:rotatebosons}) we find
\begin{equation}
  \label{eq:radii}
  R_{\rm c} N_{\rm c}={1\over\sqrt{2(m+n)}}({\cal N}_1+{\cal N}_2),\qquad
  R_{\rm n} N_{\rm n}={1\over\sqrt{2(m-n)}}({\cal N}_2-{\cal N}_1).
\end{equation}
Since ${\cal N}_i\in \Zint$ we see that the choice
$R_{\rm c}=1/\sqrt{2(m+n)}$, $R_{\rm n}=1/\sqrt{2(m-n)}$ restricts
$N_{\rm c}={\cal 
N}_1+{\cal N}_2$, $N_{\rm n}={\cal N}_2-{\cal N}_1$ to be integers.  From
this relation we also see that $N_{\rm c}$ and $N_{\rm n}$ are not independent
since they must satisfy the ``gluing condition'' $N_{\rm c}+N_{\rm n}\in
\Zint_{{\rm even}}$ in order to describe the original ${\cal N}_1{\cal
N}_2$ integer lattice.  If one considers the electron sublattice
containing the ground state (\ie, the origin ${\cal N}_i=N_{\rm
i}=0$), then one may choose a unit cell such that the $m^2-n^2$ points
in distinct electron sublattices are labeled by
\begin{equation}
  \label{eq:unitcell}
  \{(N_{\rm c},N_{\rm n})|N_{\rm n}=0,-1,\ldots-(m-n-1); 
  N_{\rm c}-F_{N_{\rm n}}=0,2,\ldots 2(m+n-1)\},
\end{equation}
where the offset $F_{N_{\rm n}}\equiv [1-(-1)^{N_{\rm n}}]/2$ is zero
if $N_{\rm n}$ is even and one if $N_{\rm n}$ is odd.  From this it is
clear that the $m^2-n^2$ distinct sectors of the full edge theory
correspond to $f_{\rm n}\equiv m-n$ sectors for the neutral boson and
$f_{\rm c}\equiv m+n$ sectors for the charged boson.  The requirement
that the vertex operator $e^{-i\alpha\phi_{\rm i}}$ changes the
topological charge $N_{\rm i}$ by an integer restricts the allowed
values of $\alpha$ to the set $R_{\rm i}\Zint$.  In terms of their
action in the topological charge lattice, the vertex operators with
$\alpha=jR_{\rm i}$ for $j=1,2,\ldots (f_{\rm i}-1)$ take us between
the different sectors of the $\phi_{\rm i}$ field.

The partition function and correlation functions can readily be
obtained since ${\cal H}_0$ is quadratic.  If we imagine that our
system is connected to a reservoir (\eg, a normal metal) such that the
edge can gain or lose electrons without disturbing the bulk state,
then the physically meaningful partition function is obtained by
summing only over states in a single sector.  Going between sectors
requires the addition of an edge quasiparticle, which must be
accompanied by the creation of a bulk quasihole, for which there is a
finite energy gap.  At finite length $L$ and inverse temperature
$\beta$ the partition function in the ground state sector (\ie, the
sector containing $|0,0\rangle$) can be written
\beqarr
  Z_{0}(\beta)=Tr(e^{-\beta {\cal H}_0})&=&
  \chi_{R_{\rm c}}^{(2(m+n),0)}(q_{\rm c})
  \chi_{R_{\rm n}}^{(2(m-n),0)}(q_{\rm n}) \nonumber \\
  &&{}+\chi_{R_{\rm c}}^{(2(m+n),m+n)}(q_{\rm c})
\chi_{R_{\rm n}}^{(2(m-n),m-n)}(q_{\rm n}), 
  \label{eq:Zw/otun}
\eeqarr
where $q_{\rm i}\equiv e^{-2\pi\beta v_{\rm i}/L}$, and
\begin{equation}
  \label{eq:CBchar}
  \chi_{R}^{(r,s)}(q)\equiv {1\over \varphi(q)}\sum_{p\in\Zint}
  q^{{1\over 2}[R(rp+s)]^2},\quad{\rm with}\quad
  \varphi(q)\equiv \prod_{k=1}^{\infty}(1-q^k),
\end{equation}
The function $\chi_{R}^{(r,s)}(q)$ is the character of a radius $R$
compact chiral boson whose topological charge is restricted to the set
$r\Zint + s$.  We note in passing the well known result that
$\chi_{1}^{(1,0)}(q)$ is equivalent to the character of a chiral Dirac
fermion on a circle with antiperiodic boundary conditions [\ie,
$\psi(x+L)=-\psi(x)$]\cite{stone,vondelft}.

When writing correlation functions we work at zero temperature, in
imaginary time $\tau=it$, and in the limit $L\rightarrow \infty$,
$a\rightarrow 0$.  Since this involves evaluating ground state
expectation values of operators that preserve the parity of $N_{\rm c}
+N_{\rm n}$, the gluing condition does not complicate the calculation.
We choose a non-standard engineering dimension for the electron
operators so that their correlation functions are well behaved in the
limit $L\rightarrow \infty$,
\begin{equation}
  \label{eq:elec_ops}
  \Psi_i(x)={1\over L^{m/2}} e^{i u_i(x)}.
\end{equation}
The electron propagator is
\begin{eqnarray}
  -{\cal G}_{ij}(\tau,x)&\equiv&\langle
  T_\tau
  \,:\!\Psi_i(\tau,x)\!\!:\,
  \,:\!\Psi_j^\dagger(0,0)\!:\,
  \rangle\nonumber \\
  &=&{\delta_{ij}\over{\left[{\X {\rm c}}\right]^{(m+n)\over 2}\left[{\X
            n}\right]^{(m-n)\over 2}}}.
  \label{eq:Gw/otun}
\end{eqnarray}
The equal-time Green's function has a spatial dependence ${\cal
G}_{ii}(0,x)\sim x^{-m}$ which clearly signals Luttinger liquid
behavior for $m\neq 1$.  The two-point function of the density
operator (\ref{eq:rho,Q}) is:
\begin{eqnarray}
  {\cal D}_{ij}(\tau,x)&\equiv&\langle
  T_{\tau}\rho_i(\tau,x)\rho_j(0)\rangle\nonumber \\ &=&{1\over
    2(m+n)}{1\over \left[{\X {\rm c}}\right]^2}+ {(-1)^{i+j}\over 2(m-n)}
    {1\over \left[{\X {\rm n}}\right]^2}.
  \label{eq:Dw/otun}
\end{eqnarray}
Note that the off-diagonal term, ${\cal D}_{12}={\cal D}_{21}$, is
non-zero even in the absence of tunneling; it vanishes only when
there are no interlayer correlations in the bulk ($n=0$) and no
interlayer interactions at the edge ($g=0$).

\subsection{Interlayer Tunneling and the Quantum Chiral sine-Gordon Equation}

We now consider adding interlayer electron tunneling to the edge
theory.  Interlayer quasiparticle tunneling at the edge is not
considered since the addition of a quasiparticle to the edge requires
the creation of a bulk quasihole, and hence it is not a low-energy
excitation.  The electron operators defined above have the property
that
\begin{equation}
  \label{eq:vertex_CR}
  e^{i u_i(x)}e^{i u_j(x')}=e^{i u_j(x')}
  e^{i u_i(x)}e^{i\pi \,{\rm sgn}(x'-x) K_{ij}},\qquad
  \hbox{for}\qquad x \neq x'. 
\end{equation}
If $K_{12}=n$ is even, this implies that the electron operators in
different layers commute rather than anticommute.  This can be
remedied by the introduction of Klein factors constructed from the
topological charges in such a way that the tunneling Hamiltonian
defined below is not modified.  The details of this procedure are
given in Appendix \ref{sec:Kleins}. 

The Hamiltonian describing interlayer single-electron tunneling is:
\begin{equation}
  \label{eq:H_1}
  {\cal H}_{1}=\lambda_0 \ix \left[\,:\!\Psi_1(x)\, {\Psi_2}^\dagger (x)\!:\, 
  +\,\,{\rm h.c.}\right],
\end{equation}
where $\lambda_0$ is the tunneling amplitude, which is assumed to be
uniform along the edge.  Using the transformation
(\ref{eq:rotatebosons}) and removing the normal ordering with
Eq.~(\ref{eq:normal-ordering}), the full Hamiltonian ${\cal H}={\cal
H}_0+{\cal H}_1$ can be expressed in terms of the rotated Bose fields
as
\begin{eqnarray}
  {\cal H}&=&\ix \left[{{1\over{4\pi}}v_{\rm c}
      \,:\!(\delx\phi_{\rm c})^2\!:\,
      +{1\over    4\pi}v_{\rm n}
      \,:\!(\delx\phi_{\rm n})^2\!:\,
      + {2 \lambda \over (2\pi a)^{\bh^2/2}}
      \cos(\bh\phi_{\rm n})}\right] \nonumber\\ &\equiv &{\cal
      H}_F[\phi_{\rm c}] 
      +{\cal H}_{\chi SG}^{\bh^2}[\phi_{\rm n}],\label{eq:H(phi)}
\end{eqnarray}
where $\bh\equiv \sqrt{2(m-n)}$, and $\lambda\equiv\lambda_0 L^{-n}$
has length dimension $(\bh^2/2-2)$.  Note that because of our
normalization of the bosonic fields, $\bh$ differs from the inverse period
conventionally used in the SG theory by a factor of $\sqrt{2\pi}$.

The Hamiltonian separates into two commuting pieces, a free chiral
boson Hamiltonian (${\cal H}_F$) for the field $\phi_{\rm c}$, and a
chiral sine-Gordon Hamiltonian (${\cal H}_{\chi SG}^{\bh^2}$) for the
field $\phi_{\rm n}$.  If we had taken $v\neq v'$ there would be a
cross term in the Hamiltonian of the form $\delx \phi_{\rm
c}\delx\phi_{\rm n}$, while if we had taken $m \neq m'$ the two pieces
of the Hamiltonian would not commute.  For the case of $v\neq v'$ we
could perform a further orthogonal transformation on the Bose fields
which would preserve the commutation relations~(\ref{eq:phi_CR}) and
eliminate the cross term in the Hamiltonian, but then the tunneling
term would involve both transformed fields\cite{nps-wip}.

An important property of the tunneling Hamiltonian, ${\cal H}_1$, is
that it commutes with ${\cal H}_0$.  This follows from
Eq.~(\ref{eq:comm-Verb-Ab1}) in Appendix \ref{sec:comm}.
Physically, it is a consequence of the fact that the translationally
invariant perturbation conserves momentum, which in a chiral system
with a single velocity is proportional to the energy.  It follows that
first-order degenerate perturbation theory for this system is exact,
and therefore the exact eigenvalues of ${\cal H}^{\bh^2}_{\chi SG}$
depend linearly on $\lambda$.

Since the zero-energy ground state of ${\cal H}_0$, $|0,0\rangle$, is
non-degenerate, another consequence of the vanishing commutator,
$[{\cal H}_0,{\cal H}_1]=0$, is that 
$|0,0\rangle$ is also an exact eigenstate of the tunneling
Hamiltonian.  In fact, ${\cal H}_1$ annihilates the ground state of
${\cal H}_0$, as we now show.  Using the mode expansion
(\ref{eq:phi_modes}) we can write
\beqarr  
  {1\over a^{{\bh^2}/2}}\cos\left({\bh\phi_{\rm n}(x)}\right)
  &=&{1\over 2}\left({2\pi\over 
  L}\right)^{{\bh^2}/2} e^{-i\pi\bh^2 x/L} \nonumber \\
  &&{}\times \left[{e^{-i\bh\chi_{\rm n} /R_{\rm n}}e^{i\bh\phi_{\rm n}^{(+)}(x)}
  e^{i\bh\phi_{\rm n}^{(-)}(x)}e^{2\pi i \bh R_{\rm n} N_{\rm n} x/L} + 
  (\bh\rightarrow -\bh)}\right], 
  \label{eq:sepvertex}
\eeqarr
where we have introduced
\begin{equation}
  \label{eq:annboson}
  \phi_{\rm n}^{(-)}(x)\equiv \sum_{r=1}^{\infty} {1\over \sqrt{r}}e^{-k_r
    a/2} e^{ik_r x}b_{{\rm n}r},\qquad \phi_{\rm n}^{(+)}\equiv
  (\phi_{\rm n}^{(-)})^{\dagger}.
\end{equation}
If we act on the state $|0,0\rangle$ with Eq.~(\ref{eq:sepvertex}) and
imagine expanding the exponentials involving $N_{\rm n}$ and $\phi_{\rm n}^{(-)}$
we see that only the zeroth-order term survives.  We then have
\begin{eqnarray}
\lefteqn{\ix {1\over a^{{\bh^2}/2}}\cos\left({\bh\phi_{\rm
  n}(x)}\right)|0,0\rangle = 
  \Biggl\{{{1\over 2}\left({2\pi\over
            L}\right)^{{\bh^2}/2}e^{-i\bh\chi_{\rm n}/R_{\rm n}}\ix
  e^{-i\pi\bh^2 
          x/L}}}& & \nonumber \\ &&
    \times {\Biggl[{1+\!\sum_{r=1}^{\infty}
            \sum_{s_1=1}^{\infty}\!\cdots\!\sum_{s_r=1}^{\infty}  
            {e^{(-2\pi ix/L)\sum_{j=1}^{r}s_j}}\,{(i\bh)^r e^{(-\pi
                a/L)\sum_{j=1}^{r}s_j} \over r!\sqrt{\prod_{j=1}^{r}s_j}}
            \prod_{j=1}^{r}b_{{\rm
  n}s_j}^{\dagger}}\Biggr]+(\bh\rightarrow -\bh) 
        }\Biggr\}|0,0\rangle, 
  \label{eq:cosineongs}
\end{eqnarray}
where we have expanded the exponential of $\phi_{\rm n}^{(+)}$.  The $x$
integral vanishes in every term (since $\bh^2 \in \Zint_{\rm even}^{+}$),
establishing that $|0,0\rangle$ is an exact zero-energy eigenstate of
${\cal H}_0+{\cal H}_1$.  Of course this does not imply that this
state is the ground state of the full Hamiltonian. 

The chiral sine-Gordon Hamiltonian we have arrived at from the bilayer
quantum Hall system has the properties: $\bh^2\in \Zint_{\rm
even}^{+}$, and $R_{\rm n}=1/\bh$.  It is natural to ask if the \XSG theory
can be sensibly defined for more general choices of these parameters.
As discussed above, the requirement that $\cos(\bh\phi_{\rm n})$ changes the
winding number by an integer gives the restriction $\bh/R_{\rm n}=p$ for
some $p\in\Zint$.  Next consider the behavior of the perturbation
under $x\rightarrow x+L$,
\beq
\label{eq:period}
\cos \bh\phi_{\rm n}(x+L)={1\over 2}e^{-i\pi \bh^2}\left({
e^{i\bh\phi_{\rm n}(x)}e^{2\pi i \bh R_{\rm n} N_{\rm n}}+
e^{-i\bh\phi_{\rm n}(x)}e^{-2\pi 
i\bh R_{\rm n} N_{\rm n}}}\right).
\eeq 
Since this operator appears in the Hamiltonian we demand that the
periodicity be independent of the state on which it acts, which
implies $\bh R_{\rm n}=p'$ for some integer $p'$.  Combining these two
conditions yields $\bh^2=pp'$, and $R_{\rm n}^2=p'/p$.  Without loss of
generality we can take $p$ and $p'$ to be positive.  We find that in
addition to the case $\bh^2\in\Zint_{\rm even}^{+}$ previously
considered, there is the case $\bh^2\in\Zint_{\rm odd}^{+}$.  We shall
refer to these as the ``bosonic'' and ``fermionic'' sequences,
respectively, since from Eq.~(\ref{eq:period}) the perturbation is
periodic for even $\bh^2$ and antiperiodic for odd $\bh^2$.  

To reiterate, for non-integer values of $\bh^2$ there does not exist a
choice of the radius $R_{\rm n}$ that ensures that the Hamiltonian
density changes the topological charge by an integer and has a
periodicity under $x\rightarrow x+L$ that is independent of the state
on which it acts.  For a given value of $\bh^2$, different choices of
the radius do not correspond to distinct theories.  If for a fixed
value of $\bh^2$ we take the minimum value of $R_{\rm n}$, we recover
the condition $R_{\rm n}=1/\bh$.  The other allowed radii at this
value of $\bh^2$ are integer multiples of this minimum radius and by
Eq.~(\ref{eq:CBchar}), they correspond to restricting the allowed
values of the topological charge in the theory with the minimum
$R_{\rm n}$.

\section{RG Flow of the Chiral sine-Gordon Theory}
\label{sec:RG}

At the end of the previous section we found that the \XSG theory
appears to be sensibly defined only at discrete values of $\bh^2$.
This suggests that $\bh$ does not flow under a renormalization group
transformation.  In this section we examine this question directly.
We work in the limit $L\rightarrow \infty$ and consider the Euclidean
action of the \XSG theory \cite{jackiw}, 
\begin{equation}
  \label{eq:XSGaction}
  S_{\chi SG}[\phi]=\int dx\,d\tau\,\left({{1\over
        4\pi}\left[{i\delt\phi\,\delx\phi+(\delx\phi)^2}\right]
        +{\lh\over (2\pi 
        a)^2}\cos(\bh\phi)}\right),
\end{equation}
where $\lh\equiv 2\lambda(2\pi a)^{2-\bh^2/2}$ is the dimensionless
coupling and in this section we set $v_{\rm n}=1$ for simplicity.
Here and in the remainder of the paper we will suppress the subscript
on the neutral boson, \ie, $\phi\equiv \phi_{\rm n}$.  Recall that the
action of the standard (\ie, non-chiral) sine-Gordon theory
is\cite{coleman}:
\begin{equation}
  \label{eq:SGaction}
  S_{SG}[\phi]=\int dx\,d\tau\,\left({{1\over
        4\pi}\left[{(\delt\phi)^2+(\delx\phi)^2}\right] +{\lh\over
        a^2}\cos(\bh\phi)}\right).
\end{equation}
The RG behavior of $S_{SG}$ is the well known Kosterlitz-Thouless flow
\cite{kogut,wiegmann,ohta,ichinose}:
\begin{equation}
  \label{eq:SGflow}
  {d\lh\over d\ell}=\left({2-{\bh^2\over 2}}\right)\lh +{\cal O}(\lh^2),
  \qquad{d\bh^2\over d\ell}=-c_0\lh^2+{\cal O}(\lh^3),
\end{equation}
where the UV momentum cutoff decreases as $\ell$ increases, and the
value of the constant $c_0>0$ is non-universal.  In Ref.~\cite{renn},
Renn argues that $S_{\chi SG}$ can be mapped into $S_{SG}$ by a
coordinate transformation, which would imply that the RG flow of the
\XSG theory is also given by Eq.~(\ref{eq:SGflow}).  This argument is
flawed since at the level of the free action, $\lh=0$, the two
theories clearly have different numbers of degrees of freedom (the
non-chiral boson can be written as the sum of two chiral bosons) and
hence cannot be related by a change of coordinates.  The classical
(real-time) equation of motion for a free chiral boson which follows
from the Minkowski version of the quadratic part of
Eq.~(\ref{eq:XSGaction}) is $(\delx\delT+\delx^2)\phi=0$, which has a
general solution $\phi(t,x)=\phi_R(x-t)+\vartheta(t)$ where
$\vartheta(t)$ is a gauge degree of freedom.  The coordinate
transformation Renn proposed amounts to promoting this unphysical
degree of freedom into a left-moving boson, which when combined with
the physical right-moving degree of freedom, $\phi_R$, gives a
non-chiral boson.

A standard perturbative, momentum-shell RG analysis for $S_{\chi SG}$
gives to first order in $\lh$
\begin{equation}
  \label{eq:XSGflow}
  {d\lh\over d\ell}=\left({2-{\bh^2\over
        2}}\right)\lh,\qquad{d\bh^2\over d\ell}=0.
\end{equation}
We see that tunneling is irrelevant for $\bh^2>4$ and relevant for
$\bh^2<4$.  The irrelevant region includes all fractional quantum Hall
states without interlayer correlations, \ie, $n=0$ and $m=3,5,\ldots$
Whether an ${\cal O}(\lh^2)$ term in the flow of $\bh^2$ is present
(as it is for the non-chiral theory) is a difficult question to answer
unambiguously using a Wilsonian RG.  Indeed, to rigorously obtain the
$-c_0\lh^2$ term in Eq.~(\ref{eq:SGflow}) from a momentum-shell RG one
must use a smooth cutoff \cite{kogut}.  Only because of the
Euclidean (rotational) symmetry of the free part of $S_{SG}$ can this term
be found simply by using an ad-hoc regularization of the divergent
integral that would be rendered finite if one employed smooth
momentum-space slicing\cite{kogut,wiegmann,ohta,ichinose}.  The free
chiral boson action does not possess this symmetry, and even
continuous regularizations lead to different results depending on
the details of the particular cutoff procedure.

Nevertheless, we now argue that there is no flow in $\bh$ for the
\XSG theory, \ie, $d\bh^2/d\ell=0$ to all orders in $\lh$.  Consider 
calculating the correlation functions of some vertex operator
perturbatively in $\lh$, for example the two-point function:
\begin{eqnarray}
  \mu^{\alpha^2}\left\langle 
  :\!e^{i\alpha \phi(z_1)}\!:
  \,:\!e^{-i\alpha \phi(z_2)}\!:
  \right\rangle&=&
  {1\over(z_1-z_2)^{\alpha^2}}\Bigglb\{{{1+{\lh^2\over 8 a^{4-\bh^2}}
      \int{
        d^2w_1 \,d^2w_2 \,{1\over (w_1-w_2)^{\bh^2}}}}} \nonumber \\ 
  \hskip-0.3in& &
  \times {{\Biggl[{\Bigglb({(z_1 -w_1)(z_2 -w_2)\over 
              (z_1-w_2)(z_2-w_1)}\Biggrb)^{\alpha \bh}
          +(w_1 \leftrightarrow w_2)-2}\Biggr]+{\cal
  O}(\lh^4)}}\Biggrb\}, \label{eq:multi} 
\end{eqnarray}
where $z\equiv \tau-ix$, and $\mu$ is an infrared momentum cutoff
needed since we are working at $L\rightarrow \infty$.  We see that for
general $\alpha$ the multi-valuedness of the correlation function is
altered by the perturbative corrections.  This should be contrasted
with the standard SG theory, where $(w-z)$ in Eq.~(\ref{eq:multi}) is
replaced by $|w-z|$ and correlation functions of all vertex operators
are single-valued at every order in $\lh$ for arbitrary $\bh$.  Note
however that for the set of allowed vertex operators defined in
Section \ref{sec:edgetheory}, $\alpha=p/\bh$ where $p\in \Zint$, we
see from Eq.~(\ref{eq:multi}) that the ${\cal O}(\lh^2)$ term has the
same multi-valuedness as a function of $(z-z')$ as the leading-order
term.  This property holds for $N$-point correlation functions ($N>2$)
and for higher-order corrections in $\lh$, since the factor of $\bh$
cancels whenever we contract an external vertex operator [$e^{i
(p/\bh) \phi(z_i)}$] with an internal one [$e^{i \bh \phi(w_i)}$].  It
is conceivable that the theory could be defined at non-integer values
of $\bh^2$ for this subset of operators, but we still contend that
there is no flow in $\bh$.

Suppose we begin with some bare values of the parameters, $\lh_0$ and
$\bh_0$, defined for a theory with a UV momentum cutoff $\Lambda \sim
1/a$ and imagine running the RG so that the cutoff is reduced to
$\Lambda/s$ where $s>1$, and $\lh_0\rightarrow \lh(s)$,
$\bh_0\rightarrow \bh(s)$.  We allow for a field renormalization by
defining $\phi'(z')\equiv Z_{\phi}^{-1}(s)\phi(z)$, where $z'=z/s$.
The covariance of correlation functions under the RG flow implies that
for $|z'_1-z'_2|\gg (\Lambda/s)^{-1}$,
\beq
\label{eq:covar}
\left\langle :\!e^{i\alpha \phi(z_1)}\!: \,:\!e^{-i\alpha \phi(z_2)}\!: 
\right\rangle_{\bh_0}=\left\langle :\!e^{i\alpha Z_{\phi}(s)\phi'(z'_1)}\!: 
\,:\!e^{-i\alpha Z_{\phi}(s)\phi'(z'_2)}\!: \right\rangle_{\bh(s)},
\eeq
where the l.h.s.\ is evaluated in the bare theory and the r.h.s.\ in
the renormalized theory.  A key point is that any flow in $\bh$ is
governed by the field renormalization: $\bh(s)=Z_{\phi}(s)\bh_0$.  If
we choose $\alpha=p/\bh_0$ for some integer $p$, so that the
perturbative corrections on the l.h.s.\ of Eq.~(\ref{eq:covar}) do not
change the multi-valuedness of the leading-order term, then the
requirement that the same thing be true on the r.h.s., along with the
relations $Z_{\phi}(1)=1$, $\bh(1)=\bh_0$, implies $\alpha
Z_{\phi}(s)=p/\bh(s)$.  Eliminating $Z_{\phi}(s)$ and $\alpha$ from
this equation yields $\bh(s)=\bh_0$, \ie, $\bh$ does not flow.  This
argument for the non-renormalization of $\bh$ can at best be viewed as
heuristic; its validity is intricately connected to the way the RG
transformation is defined.  We believe that at least the weaker
statement that there exists an RG procedure for which there is no flow
in $\bh$, is true.  The exact solutions of the theory at $\bh^2=2$ and
$\bh^2=4$ presented in the next section support this claim.

\begin{figure}[htbp]
  \begin{center}
    \leavevmode
    \epsfxsize=0.8\columnwidth
    \epsfbox{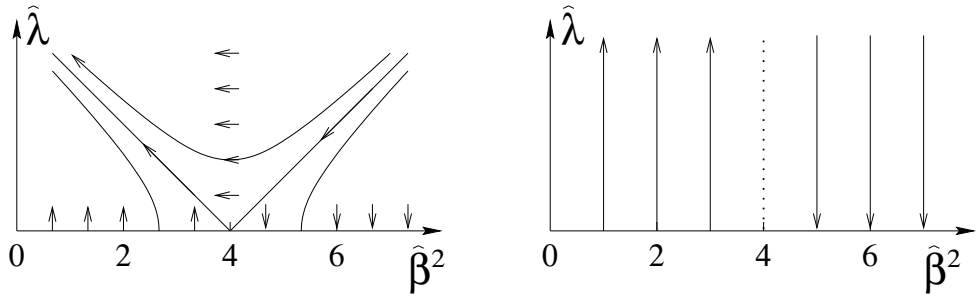}\vskip1mm
    \caption{Left: the perturbative RG flow for the standard SG
      theory.  Both $\lh$ and $\bh^2$ are renormalized.  Right: the
      perturbative RG flow for the \XSG theory.  $\bh$ does not
      flow and there is a line of fixed points at $\bh^2=4$ (dots). }
    \label{fig:flows}
  \end{center}
\end{figure}

In light of the argument that $d\bh/d\ell=0$, the flow in $\lh$ at the
lowest-order marginal point, $\bh^2=4$, will be governed by the
higher-order terms in the first expression in Eq.~(\ref{eq:XSGflow}).
Because of the complexity of the higher-order bosonic perturbation
theory, we shall postpone this calculation until we find a fermionic
representation of the theory in the next section.  We will find that at
this value of $\bh$ the tunneling perturbation is exactly marginal,
\ie, there is a line of fixed points parameterized by $\lh$.  The RG
flows of the standard SG and \XSG theories are shown in Fig.~\ref{fig:flows}.

\section{Exact Solutions of the Chiral sine-Gordon Theory}
\label{sec:exactsolns}

In this section we present exact solutions of the \XSG theory at the
points $\bh^2=2$ and $\bh^2=4$. 

\subsection{Relevant Tunneling, $\bh^2=2$}

At the point $\bh^2=2$, which corresponds to $m-n=1$, the cosine term
is relevant.  There is an infinite sequence of $K$ matrices describing
maximally chiral edges which satisfy this condition:
$(m,m',n)=(1,1,0)+r(2,2,2)$, $r=0,1,\ldots$ The discussion here also
applies to the bosonic states $(m,m',n)=(2,2,1)+r(2,2,2)$.

At this value of $\bh$ the \XSG Hamiltonian is
\begin{equation}
  \label{eq:HXSG_4pi}
  {\cal H}_{\chi SG}^{\bh^2=2}= \ix\left[{{1\over 4\pi}v_{\rm n}
      \,:\!(\delx\phi)^2\!:\,
      +{\lambda\over 2\pi a}\left(e^{i\sqrt{2}\phi}+\,{\rm
      h.c.}\right)}\right], 
\end{equation}
where $\phi(x)$ is a radius $R_{\rm n}=1/\sqrt{2}$ chiral boson.  A
similar problem, with $\lambda$ replaced by a Gaussian random
variable, appears in Kane, Fisher, and Polchinski's work on the
disordered, single-layer, $\nu=2/3$ edge\cite{kfp}.  A special
property of a radius $1/\sqrt{2}$ free chiral boson is that its
character~(\ref{eq:CBchar}) can be expressed as:
\begin{equation}
  \label{eq:su(2)char}
  \chi_{1/\sqrt{2}}^{(1,0)}(q)
  =\chi_{su(2)_1}^{(0)}(q)+\chi_{su(2)_1}^{(1/2)}(q), 
\end{equation}
where $\chi_{su(2)_1}^{(j)}(q)$ is the character of the spin-$j$
irreducible representation of the ${\widehat{su}(2)}_1$ Kac-Moody (KM)
algebra\cite{BYB}.  Indeed, the trio of dimension one operators
$\delx\phi, e^{i\sqrt{2}\phi},e^{-i\sqrt{2}\phi}$ are related to the
${\widehat{su}(2)}_1$ generators, $J^a(x)$, $a=x,y,z$, by\cite{BYB} 
\begin{equation}
  \label{eq:boson_su(2)curr}
  J^z(x)={1\over 2\pi\sqrt{2}}\delx\phi(x),\qquad J^{\pm}(x)=J^x \pm i J^y=
  {1\over 2\pi a}e^{\pm i \sqrt{2}\phi(x)}. 
\end{equation}
In terms of the Fourier components
\begin{equation}
  \label{eq:su(2)modes}
  J^a_n\equiv \ix J^a(x)e^{-2\pi i n x/L},\quad n\in\Zint,
\end{equation}
the algebra is
\begin{equation}
  \label{eq:su(2)_1KM}
  [J^a_n, J^b_m]={n\over 2}\delta^{ab}\delta_{n+m,0}+i\epsilon^{abc}
  J^c_{n+m}.  
\end{equation}
With the help of Eq.~(\ref{eq:boson_su(2)curr}) the
Hamiltonian~(\ref{eq:HXSG_4pi}) can be expressed in terms of currents
\begin{equation}
  \label{eq:H_su(2).1}
  {\cal H}_{\chi SG}^{\bh^2=2} =\ix\left\{{2\pi
      v_{\rm n}
      \,:\!\left[{J^z(x)}\right]^2\!:\,
       +2\lambda J^x(x)}\right\}.
\end{equation}
We now use an identity valid for any state $|\gamma\rangle$ in the
Hilbert space 
\begin{equation}
  \label{eq:Jz_to_J^2}
  \ix
  \,:\!\left[{J^z(x)}\right]^2\!:\,
  |\gamma\rangle=\ix{1\over 3}
  \,:\!
  \left[{{\bf J}(x)}\right]^2\!:\,
  |\gamma\rangle,
\end{equation}
to write Eq.~(\ref{eq:H_su(2).1}) as:
\begin{equation}
  \label{eq:H_su(2).2}
  {\cal H}_{\chi SG}^{\bh^2=2}=\ix\left\{{{2\pi v_{\rm n}\over 3}:
      \left[{{\bf J}(x)}\right]^2:+2\lambda J^x(x)}\right\}.
\end{equation}
One way to prove Eq.~(\ref{eq:Jz_to_J^2}) is to note that the
operators $J^z_n$ obey a $\hat{u}(1)$ subalgebra of the
$\widehat{su}(2)_1$ algebra generated by the $J^a_n$ operators.  We can
construct two families of operators
\begin{equation}
  \label{eq:su(2)_1virasoro}
  L_n\equiv {1\over 3}:\!J^a_m J^a_{n-m}\!:, \qquad L^{(z)}_n\equiv\,
  :\!J^z_m  J^z_{n-m}\!:, 
\end{equation}
each of which obey $c=1$ Virasoro algebras.  Since the $J^z_n$
generate a subalgebra we have
$[L_n,\,L^{(z)}_m]=[L^{(z)}_n,\,L^{(z)}_m]$, and therefore by the
standard Goddard-Kent-Olive (GKO) coset construction\cite{GKO} the operators
$L^{(0)}_n\equiv L_n-L^{(z)}_n$ obey a $c=0$ Virasoro
algebra.  It is well known that the only unitary
representation of the $c=0$ Virasoro algebra is trivial, \ie,
$L^{(0)}_n|\gamma\rangle=0$ for all $n$\cite{Ginsparg}.  We thus conclude that
$L_n|\gamma\rangle=L^{(z)}_n|\gamma\rangle$ for any state in the
Hilbert space, and for $n=0$ we obtain Eq.~(\ref{eq:Jz_to_J^2}).

Given any matrix $R\in SO(3)$, we see from Eq.~(\ref{eq:su(2)_1KM})
that the currents ${\tilde J}^a(x)\equiv R^{ab}J^b(x)$ also obey an
$\widehat{su}(2)_1$ KM algebra since $\delta^{ab}$ and $\epsilon^{abc}$
are $SO(3)$-invariant tensors.  If we make the particular choice
$R^{ab}=\delta^{ay}\delta^{by}+\delta^{az}\delta^{bx}
-\delta^{ax}\delta^{bz}$, which corresponds to a rotation by $\pi/2$
about the $y$-axis, and express the rotated $\widehat{su}(2)_1$ currents
in terms of a new radius $R=1/\sqrt{2}$ chiral boson, $\theta(x)$
(\ie, ${\tilde J}^z(x)=[\delx \theta(x)]/(2 \pi \sqrt{2})$, etc.) then from
Eqns.~(\ref{eq:Jz_to_J^2}) and (\ref{eq:H_su(2).2}) we have
\begin{eqnarray}
  {\cal H}_{\chi SG}^{\bh^2=2}&=& \ix\left\{{2\pi
      v_{\rm n}:\!\left[{{\tilde J}^z(x)}\right]^2\!: 
      +2\lambda {\tilde J}^z(x)}\right\} \nonumber \\
  &=&\ix\left[{{1\over 4\pi}v_{\rm n}:\!(\delx\theta)^2\!:
      +{\sqrt{2}\lambda\over 2\pi}\delx\theta}\right]. 
  \label{eq:H_4pi_final}
\end{eqnarray}

The mapping between Eqns.~(\ref{eq:HXSG_4pi}) and (\ref{eq:H_4pi_final}),
when expressed in terms of the two Bose fields corresponds to the
non-linear transformation
\begin{equation}
  \label{eq:bose-bose}
  \delx\theta(x) ={\sqrt{2}\over a}\cos\biglb(\sqrt{2}\phi(x)\bigrb),\qquad
  \delx\phi(x)= 
  -{\sqrt{2}\over a}\cos\biglb(\sqrt{2}\theta(x)\bigrb).
\end{equation}
While we could have taken the first expression in this equation as our
definition of $\theta(x)$ and proceeded by showing this implies
$:\!(\delx \phi)^2\!:\, =\,:\!(\delx \theta)^2\!:$, the detour through
the KM algebras is more deductive.

We have succeeded in writing the Hamiltonian as a quadratic form in
terms of a single chiral boson.  The tunneling term is linear in the
topological charge of the field $\theta(x)$, and thus it is clear that
the $\lambda=0$ ground state, which is annihilated by the
perturbation, is not the ground state for $\lambda\neq 0$.  However,
since the $\lambda=0$ ground state remains an exact eigenstate of
Eq.~(\ref{eq:H_4pi_final}) for all $\lambda$ [see
Eq.~(\ref{eq:cosineongs})] we conclude that zero-temperature
perturbation theory in $\lambda$ gives incorrect results.  In terms of
first-order degenerate perturbation theory for the low-lying
eigenvalues of ${\cal H}_{\chi SG}^{\bh^2=2}$, this is a consequence
of a level crossing for $\lambda = {\cal O}(1/L)$.

\begin{figure}[htbp]
  \begin{center} 
    \leavevmode 
    \epsfxsize=0.5\columnwidth
    \epsfbox{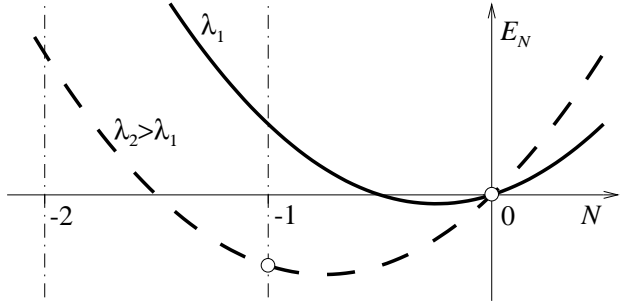} \vskip1mm
    \caption{Energy spectrum for the topological charge ($N$) of the field
    $\theta(x)$ at two different values of $\lambda$.  The open
    circles indicate the ground state.  For small $\lambda$ (solid line) 
    the ground state is unchanged, while at a larger value of $\lambda$
    (dashed line) the ground state has a non-zero winding number.}
    \label{fig:roottwo}
  \end{center}
\end{figure}

Several authors have claimed that a gap is generated when the
perturbation is relevant \cite{renn,nomura}.  It is clear from
Eq.~(\ref{eq:H_4pi_final}) that this is not the case.  We still have a
(gapless) chiral boson, whose topological charge sectors are
reorganized in energy.  The energy spectrum of the winding number mode
for two different values of $\lambda$ is shown in
Fig.~\ref{fig:roottwo}. For the points $\lambda= (\pi v_{\rm n}/L)M$,
where $M\in \Zint$, this reorganization can be reduced to an overall
shift in the energy spectrum by defining ${\tilde N}_{\theta} =
N_{\theta}+M$.  The subtlety is that while we have a gapless mode for
$\lambda=0$ ($\phi(x)$) and a gapless mode for $\lambda\neq 0$
($\theta(x)$), the modes are non-trivially connected even for
arbitrarily small $\lambda$.  We see from the relation,
Eq.~(\ref{eq:bose-bose}), that exciting modes of the original boson
involves the operator $\delx\phi\sim \cos(\sqrt{2}\theta)$, which
changes $N_\theta$ and hence involves an energy of order $\lambda$
even for small momentum excitations.  Therefore the original field
appears massive, whereas the number of gapless modes is unchanged.

From Eq.~(\ref{eq:H_4pi_final}) we can immediately write down the
partition function for the \XSG theory in isolation
\begin{equation}
  \label{eq:Z_XSG_4pi}
  Z^{\bh^2=2}_{\chi
  SG}(\beta)={1\over\varphi(q_{\rm n})}\sum_{N\in\Zint}e^{-\beta\lambda
  N}q_{\rm n}^{N^2/4}.
\end{equation}
To construct the partition function of the full edge theory we need to
consider gluing conditions.  Recall from Section \ref{sec:edgetheory}
that we need to separate states according to their $N_{\rm n}$-parity,
as in the partition function at $\lambda=0$ (\ref{eq:Zw/otun}).  For
$\lambda\neq 0$, eigenstates of the Hamiltonian are no longer
eigenstates of $N_{\rm n}$.  However, we see from
Eqns.~(\ref{eq:boson_su(2)curr}) and (\ref{eq:su(2)modes}), that
$N_{\rm n}=2J^z_0$ and similarly $N_{\theta}=2{\tilde J}^z_0$, where
$N_{\theta}$ is the winding number operator of the field $\theta(x)$.
Therefore, states in the spin-0 (1/2) representation of the $J^a_n$
algebra have $N_{\rm n}$ even (odd).  Under the rotation from $J^a$ to
${\tilde J}^a$ the two representations do not mix, and hence
eigenstates of Eq.~(\ref{eq:H_4pi_final}) which have $N_{\theta}$ even
(odd) are linear combinations of $N_{\rm n}$ eigenstates, all of which
have even (odd) $N_{\rm n}$ eigenvalues.  Thus we can write the
partition function including the charged mode as:
\begin{equation}
\label{eq:Z_4pi}
Z^{\bh^2=2}(\beta)=\chi_{R_{\rm c}}^{(2(m+n),0)}(q_{\rm c})Z_{\chi
  SG}^{\bh^2=2,\,{\rm even}}(\beta)  
+\chi_{R_{\rm c}}^{(2(m+n),m+n)}(q_{\rm c})Z_{\chi SG}^{\bh^2=2,\,{\rm
  odd}}(\beta), 
\end{equation}
where $Z_{\chi SG}^{\bh^2=2,\,{\rm even}}$ is given by
Eq.~(\ref{eq:Z_XSG_4pi}) 
with the restriction $N\in\Zint_{\rm even}$, and similarly for
$Z_{\chi SG}^{\bh^2=2,\,{\rm odd}}$.

We now turn to the computation of correlation functions.  We define
the tunneling operator as the Hamiltonian density of the tunneling
perturbation
\begin{equation}
  \label{eq:tunnelingop}
  \Theta(x)\equiv {2\over (2\pi
    a)^{\bh^2/2}}\cos\left[{\bh\phi(x)}\right],
\end{equation}
and denote its two-point correlation function by ${\cal
T}(\tau,x)\equiv\langle T_{\tau}\Theta(\tau,x)\Theta(0)\rangle$.
Using transformations (\ref{eq:rotatebosons}) and (\ref{eq:bose-bose}) we
can express the charge density (\ref{eq:rho,Q}), and tunneling
operator (\ref{eq:tunnelingop}), in terms of the fields $\phi_{\rm c}$
and $\theta$,
\begin{eqnarray}
  \rho_{1,2}(x)&=&{1\over\sqrt{8\pi^2 (2m-1)}}\delx\phi_{\rm c}(x)\pm
  {1\over 2\pi a} 
  \cos\sqrt{2}\theta(x), \label{eq:rho_free}  \\
  \Theta(x)&=&{1\over \pi \sqrt{2}}\delx\theta(x). \label{eq:Theta_free}
\end{eqnarray}
Since the Hamiltonian ${\cal H}_F[\phi_{\rm c}]+{\cal H}_{\chi
  SG}^{\bh^2=2}[\theta]$ is quadratic in these fields we find
\begin{eqnarray}
  {\cal D}_{ij}(\tau,x)&=&{1\over 2(2m-1)} {1\over \left[{\X
          {\rm c}}\right]^2}+{(-1)^{i+j}\over 2}{\cos(2\lambda
          x/v_{\rm n})\over 
          \left[{\X {\rm n}}\right]^2},\label{eq:D_4pi} \\ {\cal
          T}(\tau,x)&=&{2\over 
          \left[{\X {\rm n}}\right]^2} + \left({\lambda \over\ \pi
          v_{\rm n}}\right)^2.\label{eq:T_4pi} 
\end{eqnarray}
The part of the density-density correlation function arising from the
neutral mode is spatially modulated when tunneling is
present, while the two-point correlation function of the tunneling
operator acquires a constant piece.  Both of these are a result of the 
non-zero topological charge of the $\theta$ field in the ground state.

The calculation of the electron propagator is slightly more involved.
From Eqns.~(\ref{eq:rotatebosons}) and (\ref{eq:elec_ops})
we find that for $m-n=1$ the electron operators can be written
\begin{equation}
  \Psi_{1,2}={1\over L^{m/2}} e^{i\left({\sqrt{m-{1\over2}}\phi_{\rm c}
        \mp{1\over \sqrt{2}}\phi}\right)},
\end{equation}
and therefore they cannot be readily expressed in terms of $\phi_{\rm
  c}$ and 
$\theta$.  However, from this expression it is clear that the
single-electron Green's function factorizes:
\begin{eqnarray}
  -{\cal G}_{ij}(\tau,x)&=&{1\over L^{m/2}} \left\langle e^{i\sqrt{m-{1\over
  2}}\phi_{\rm c}(\tau,x)}e^{-i\sqrt{m-{1\over
  2}}\phi_{\rm c}(0)}\right\rangle_{{\cal H}_F[\phi_{\rm c}]} \nonumber\\ 
  & & \times\left\langle e^{{i\over \sqrt{2}}(-1)^i\phi(\tau,x)}
  e^{-{i\over \sqrt{2}} 
  (-1)^j\phi(0)} \right\rangle_{{\cal H}_{\chi SG
  }^{\bh^2=2}}. 
  \label{eq:factorprop}  
\end{eqnarray}
The correlation function involving the $\phi$ field is the same for
all values of $m$ and $n$ that satisfy $m-n=1$.  The correlation
function involving $\phi_{\rm c}$ can be computed for all $m$ and $n$ since
${\cal H}_F[\phi_{\rm c}]$ is quadratic.

For the special case $m=1,n=0$ the l.h.s.\ of
Eq.~(\ref{eq:factorprop}) can be calculated exactly as we now
demonstrate\cite{Larkin}.  For the uncorrelated integer 110 state
there exists a chiral fermion description of the edge theory including
tunneling, ${\cal H}_0+{\cal H}_1$:
\begin{equation}
\label{eq:H(110)}
  {\cal H}^{(110)}=\ix:\!\left[{-iv\psi_i^{\dagger}\delx\psi_i +2\pi g
  \psi_1^{\dagger}\psi_1 \psi_2^{\dagger}\psi_2
  -\lambda(\psi_2^{\dagger}\psi_1+\,{\rm h.c.})  }\right]\!:.
\end{equation}
The details in going from the bosonic form of the Hamiltonian
(\ref{eq:H(phi)}) to the above fermionic form are discussed in
Section \ref{sec:fermionizations}.  If we define
$\psi_{\pm}(x)=\biglb(\psi_1(x) \pm \psi_2(x)\bigrb)/\sqrt{2}$,
Eq.~(\ref{eq:H(110)}) becomes
\beq
\label{eq:H+/-}
{\cal H}^{(110)}=\ix:\!\left[{-iv\left({\psi_+^{\dagger}\delx\psi_+ +
 \psi_-^{\dagger}\delx\psi_-}\right) +2\pi g
  \psi_+^{\dagger}\psi_+ \psi_-^{\dagger}\psi_-
  -\lambda(\psi_+^{\dagger}\psi_+ -\psi_-^{\dagger}\psi_-)}\right]\!:.
\eeq
This Hamiltonian can be brought to a quadratic form by bosonizing
according to $\psi_{\pm}(x)=e^{i\phi_{\pm}(x)}/\sqrt{2\pi a}$.  
The resulting quadratic form is diagonalized by defining 
$\theta_+\equiv(\phi_+ +\phi_-)/\sqrt{2}$, and 
$\theta_-\equiv(\phi_+-\phi_-)/ \sqrt{2}+{\sqrt{2}\lambda x/ v_{\rm n}}$.
The exact single-electron Green's function is then readily found:
\begin{equation}
  \label{eq:G110}
  -{\cal G}_{ij}^{(110)}(\tau,x)={{\delta_{ij}\cos(\lambda
      x/v_{\rm n})-(1-\delta_{ij})i\sin(\lambda x/v_{\rm
      n})}\over\sqrt{\X {\rm c} \X {\rm n}}}. 
\end{equation}
We can thus use Eqns.~(\ref{eq:factorprop}) and (\ref{eq:G110}) to
find the single-electron Green's function for all states in the
$m-n=1$ sequence:
\begin{equation}
  \label{eq:G_4pi}
  -{\cal G}_{ij}(\tau,x)={{\delta_{ij}\cos(\lambda
      x/v_{\rm n})-(1-\delta_{ij})i\sin(\lambda x/v_{\rm n})} \over\left[
   {\X {\rm c}}\right]^{m-{1 \over 2}}\sqrt{\X {\rm n}}},
\end{equation}
the only modification being the exponent of the piece corresponding to
the charged boson.

\subsection{Marginal Tunneling, $\bh^2=4$}

The point $\bh^2=4$, at which the cosine term is marginal, corresponds
to the case $m-n=2$.  There is again an infinite sequence of $K$
matrices describing maximally chiral edges which satisfy this
condition: $(m,m',n)=(3,3,1)+r(2,2,2)$, $r=0,1,\ldots$  The
results here also apply to the bosonic states
$(m,m',n)=(2,2,0)+r(2,2,2)$.

At this value of $\bh$ the \XSG Hamiltonian is
\begin{equation}
  \label{eq:newHXSG8pi}
  {\cal H}_{\chi SG}^{\bh^2=4}= \ix\left[{{1\over 4\pi}v_{\rm n}
      \,:\!(\delx\phi)^2\!:\,
      +{\lambda\over (2\pi a)^2}\left(e^{2i\phi}+\,{\rm
      h.c.}\right)}\right], 
\end{equation}
where $\phi(x)$ is a radius $R_{\rm n}=1/2$ chiral boson.  If we
restrict ourselves to the sector of the edge theory that contains the
ground state, then we find that $N_{\rm n}\in \Zint_{\rm even}$, since
adding an electron to layer $1(2)$ changes $N_{\rm n}$ by $\mp 2$.
From the expression for the compact chiral boson character
(\ref{eq:CBchar}), it is clear that a radius $1/2$ boson restricted to
even topological charges is equivalent to a radius $1$ boson with no
restriction [\ie, $\chi_{1/2}^{(2,0)}(q) =\chi_{1}^{(1,0)}(q)$].
Hence, for this sector of the edge theory we can take $\phi(x)$ to
have radius $1$ and thus define a chiral Dirac fermion $\psi(x)\equiv
e^{i\phi(x)}/\sqrt{2\pi a}$.  This result, that the edge theory of
double-layer states with $m-n=2$ contains a chiral Dirac fermion was
previously established in the wavefunction approach by Milovanovi\'{c}
and Read\cite{milo&read}.

To find a fermionic representation of the tunneling term, we first
consider bringing together two chiral Dirac annihilation operators
\begin{equation}
  \label{eq:ptsplitfermi}
  \psi(x)\psi(x+\epsilon)=\epsilon\psi(x)\delx\psi(x)+{\cal
    O}(\epsilon^2),
\end{equation}
where the zeroth-order term vanishes by Fermi statistics.  We next
consider bringing together the bosonized representation of the Fermi
fields
\begin{eqnarray}
  {1\over L}\,:\!e^{i\phi(x)}\!:\,:\!e^{i\phi(x+\epsilon)}\!:\, &
  =&{2\pi a\over L^2}\,:\!e^{2i\phi(x)}\!: \nonumber \\ 
  \label{eq:ptsplitbose} & & +{2\pi i \epsilon\over
    L^2}\left\{{\,:\!e^{2i\phi(x)}\!:-a\,:\!\left[{ \delx\phi(x) +
          {3\pi\over L}}\right]e^{2i\phi(x)}\!:\, }\right\}+{\cal
    O}(\epsilon^2,a^2).
\end{eqnarray}
Since the terms on the r.h.s.\ are normal ordered, we can safely take
the limit $a\rightarrow 0$, and compare the order $\epsilon$ terms in
Eqns.~(\ref{eq:ptsplitfermi}) and (\ref{eq:ptsplitbose}) to obtain the
identity
\begin{equation}
  \label{eq:fermie2phi}
  \psi(x)\delx\psi(x)={i\over 2\pi a^2}e^{2i\phi(x)}.
\end{equation}
We verified the validity of this formula by checking that the
$N$-point correlation functions of the operator on the l.h.s.\ (and
its Hermitian conjugate) calculated in the free Fermi theory are
identical to the $N$-point correlation functions of the operator on
the r.h.s.\ (and its Hermitian conjugate) calculated in the free Bose
theory.  The proof of this equivalence is given in Appendix
\ref{sec:corr}.  In Appendix \ref{sec:comm} we demonstrate that the
correspondence holds for the commutation relations as well.  Readers
untroubled by operator manipulations in bosonization can ignore them.

Using Eq.~(\ref{eq:fermie2phi}) we can then fermionize the \XSG
Hamiltonian (\ref{eq:newHXSG8pi}) to
\begin{equation}
  \label{eq:HXSG8pi,Dirac}
  {\cal H}_{\chi SG}^{\bh^2=4}=\ix
  \,:\!\left[{-iv_{\rm n}\psi^{\dagger}\delx\psi-i{\lambda\over
        2\pi}(\psi^{\dagger}
    \delx\psi^{\dagger}+\psi\delx\psi)}\right]\!:. 
\end{equation}
The effect of the tunneling term is made transparent by writing the
Dirac fermion in terms of its Majorana components: $\psi(x)\equiv
[\chi_1(x)+i\chi_2(x)]/\sqrt{2}$, where $\chi_i^{\dagger}=\chi_i$,
\begin{equation}
  \label{eq:HXSG8pi,Maj}
  {\cal H}_{\chi SG}^{\bh^2=4}=-{1\over
    2}\ix
  \,:\!\left[{i\biggl({v_{\rm n}+{\lambda\over \pi}}\biggr)\chi_1\delx\chi_1
      +i\biggl({v_{\rm n}-{\lambda\over\pi}}\biggr)\chi_2\delx\chi_2}\right]\!:.
\end{equation}
An alternative derivation of this form of the Hamiltonian is given in
Appendix \ref{sec:direct}.  We see that the effect of tunneling is to
split the neutral mode into two Majorana fields with different
velocities.  This is a $c=1=(1/2)+(1/2)$ analog of the usual
spin-charge separation in one dimension [$c=2=(1)+(1)$].  Since the
Hamiltonian (\ref{eq:HXSG8pi,Dirac}) is quadratic, modes of the
fermion field with momentum $|k|\ge \Lambda$, for some $\Lambda$, do
not couple to modes with momentum $|k|<\Lambda$, and thus the
tree-level RG flow is exact.  Therefore if we define our RG
transformation so that the $\lambda=0$ Hamiltonian is a fixed point,
we immediately see that the coupling $\lambda$ is exactly marginal,
and we have a line of fixed points as asserted at the end of Section
\ref{sec:RG}.

Armed with a quadratic representation of the Hamiltonian we can
readily compute the partition function and correlation functions
exactly.  From the expression for the partition
function~(\ref{eq:Zw/otun}) at $\lambda=0$ we see that the characters
of the neutral mode that appear at the $\bh^2=4$ point are
$\chi_{1/2}^{(4,0)}$ and $\chi_{1/2}^{(4,2)}$.  These can be written
in terms of the characters of the highest weight $h$ representation of
the Virasoro algebra with central charge $c$, denoted ${\cal
  V}^{c}_{h}$:
\begin{equation}
  \label{eq:VirChar}
  \chi_{1/2}^{(4,0)}(q)=\left[{{\cal
        V}^{1/2}_{0}(q)}\right]^2+\left[{{\cal
        V}^{1/2}_{1/2}(q)}\right]^2,\quad \chi_{1/2}^{(4,2)}(q)=2{\cal
    V}^{1/2}_{0}(q)\,{\cal V}^{1/2}_{1/2}(q).
\end{equation}
Since a single chiral Majorana fermion with antiperiodic boundary
conditions has a character ${\cal V}^{1/2}_{0}+{\cal V}^{1/2}_{1/2}$,
we see from Eqns.~(\ref{eq:Zw/otun}), (\ref{eq:HXSG8pi,Maj}), and
(\ref{eq:VirChar}), that the partition function of the \XSG theory in
isolation is
\begin{equation}
\label{eq:Z_XSG_8pi}
Z^{\bh^2=4}_{\chi SG}(\beta)=\left[{\cal V}^{1/2}_{0}(q_+)+{\cal
    V}^{1/2}_{1/2}(q_+)\right] \, \left[{\cal V}^{1/2}_{0}(q_-)+{\cal
    V}^{1/2}_{1/2}(q_-)\right], 
\end{equation}
where $q_{\pm}\equiv e^{-2\pi\beta v_{\pm}/L}$, and $v_{\pm}\equiv
v_{\rm n}\pm\lambda/\pi$, while the partition function of the full edge
theory with tunneling, including the charged mode with the appropriate
gluing condition, is
\begin{eqnarray}
  Z^{\bh^2=4}(\beta)&=&\chi_{R_{\rm c}}^{(2(m+n),0)}(q_{\rm c})\left[{{\cal
        V}^{1/2}_{0}(q_+){\cal V}^{1/2}_{0}(q_-) +{\cal
        V}^{1/2}_{1/2}(q_+){\cal V}^{1/2}_{1/2}(q_-)}\right]\nonumber\\
  & &+\chi_{R_{\rm c}}^{(2(m+n),m+n)}(q_{\rm c})\left[{{\cal V}^{1/2}_{0}(q_+){\cal
    V}^{1/2}_{1/2}(q_-)+{\cal V}^{1/2}_{1/2}(q_+){\cal
    V}^{1/2}_{0}(q_-)}\right].
    \label{eq:Z8pi}
\end{eqnarray}

Recall that there are three irreducible Verma modules in the $c=1/2$
minimal model with highest weights $h=0,1/2$, and $1/16$, which in the Ising
model terminology correspond to the identity ($\ID$), energy
($\epsilon$) and spin ($\sigma$) fields.  The characters of the first
two of these representations appear in the above partition function,
the remaining one occurs if one considers the other sector of the
neutral mode which involves the characters $\chi_{1/2}^{(4,1)}$
and $\chi_{1/2}^{(4,3)}$, and which corresponds to the addition of
a quasiparticle.  These combine to give
$\chi_{1/2}^{(4,1)}+\chi_{1/2}^{(4,3)}=2\left({{\cal
V}^{1/2}_{1/16}}\right)^2$.

Turning now to correlation functions, we can use the transformation
(\ref{eq:rotatebosons}), and the fermionization of the neutral boson to
write the electron (\ref{eq:elec_ops}), density (\ref{eq:rho,Q}), and
tunneling (\ref{eq:tunnelingop}) operators in terms of the fields
$\phi_{\rm c}$ and $\psi$,
\beqarr
:\!\Psi_i\!:\, &=& {1\over (2\pi a)^{(m-1)/2}} e^{i\sqrt{m-1}\phi_{\rm c}}
\left({\delta_{i1}\psi^{\dagger}+\delta_{i2}\psi}\right), \label{eq:elec8pi} \\
\rho_i &=& {1\over 4\pi\sqrt{m-1}}\delx\phi_{\rm c}+{(-1)^i\over 2}
\,:\!\psi^\dagger\psi\!:, \label{rho8pi} \\
\Theta(x)&=&-{i\over 2\pi}\,:\!\left({\psi^{\dagger}\delx\psi^{\dagger} 
+\psi\delx\psi}\right)\!:\,, \label{tun8pi} 
\eeqarr
where $i=1,2$ is the layer index.  The single-electron Green's
function is
\beq
 \label{eq:G_8pi}
 -{\cal G}_{ij}(\tau,x)={1\over 2\left[{\X {\rm c}}\right]^{m-1}} 
 \left[{{1\over \X +}+{(-1)^{i+j}\over \X -}}\right],
\eeq
and we see that in the presence of tunneling an electron has
broken up into three pieces, propagating with three different
velocities.  The two-point functions of the density
and tunneling operators are:
\begin{eqnarray}
  {\cal D}_{ij}(\tau,x)&=&{1\over 4(m-1)}{1\over \left[{\X {\rm c}}\right]^2}
  +{(-1)^{i+j}\over 4}{1\over \X + \,\X -}, \label{eq:D_8pi} \\ 
  {\cal T}(\tau,x)&=&{1\over \left[{\X +}\right]^4}+
  {1\over \left[{\X -}\right]^4}. \label{eq:T_8pi}
\end{eqnarray}

\section{Fermionizations of the Chiral sine-Gordon Theory}
\label{sec:fermionizations}

In this section we describe a general method for fermionizing the \XSG
theory for $\bh^2 \in \Zint^+$ by the introduction of auxiliary
degrees of freedom.  The procedure is illustrated in detail for the
cases $\bh^2=2$ and $\bh^2=4$.  We also show that for these two points
the fermionized theory with auxiliary degrees of freedom can also be
solved exactly, and the projection onto the \XSG Hilbert space can be
performed explicitly, reproducing the results of the exact solutions
given above.  In terms of the solution of the double-layer
problem, the method described here is redundant.  However, we believe
it may be useful in the study of quantum Hall systems with more than
two layers.  This section can be skipped since the subsequent sections
are independent of the material presented here.

\subsection{General Strategy}

In the case of the standard SG theory it is possible to fermionize the
theory at all values of $\bh^2 \leq 4$ to the massive Thirring
model\cite{coleman}.  For the \XSG theory we have found fermionizations
at the discrete points $\bh^2 \in \Zint^{+}$.  Recall that all the
points in the bosonic sequence ($\bh^2 \in \Zint_{{\rm even}}^{+}$)
correspond to some double-layer quantum Hall system, while none of the
points in the fermionic sequence ($\bh^2 \in \Zint_{{\rm odd}}^{+}$)
do.  The scaling dimension of the tunneling perturbation is an integer
in the bosonic sequence and a half-integer in the fermionic one.  We
defer discussion of the fermionic sequence until Section
\ref{sec:fermiseq}.

The algebraic identity at the heart of fermionization (or
bosonization) is the Cauchy formula for the determinant of an $N\times
N$ matrix $M^{(N)}_{ij}\equiv 1/(z_i-z'_j)$ which is a function of
$2N$ complex variables $z_i,z'_i, i=1,\ldots,N$ \cite{BYB}:
\begin{equation}
  \label{eq:cauchy}
  (-1)^{N(N-1)/2}\det M^{(N)}={\prod_{1\leq i < j\leq N} (z_i-z_j)
    (z'_i-z'_j) \over \prod_{1\leq i,j \leq N}(z_i-z'_j)}.
\end{equation}
The determinant on the l.h.s.\ is proportional to the $2N$-point
correlation function of a single chiral Dirac fermion:
\begin{equation}
  \label{eq:2N-Dirac}
  \left\langle{\prod_{i=1}^{N}
  \psi(\tau_i,x_i)\prod_{i=1}^{N}\psi^{\dagger}(\tau'_i,x'_i)}\right\rangle 
  ={(-1)^{N(N-1)/2}\over (2\pi)^N}\det M^{(N)},
\end{equation}
where $z_i\equiv (v\tau_i-i x_i)$, and similarly for the primed
coordinates, $v$ is the velocity of the fermion, and time ordering
will be implicit from here on.  We desire a fermionic representation
of the tunneling operator, Eq.~(\ref{eq:tunnelingop}).  The $2N$-point
correlation function of the tunneling operator is a linear combination
of terms of the form
\begin{equation}
  \label{eq:2N-tunneling}
  {1\over a^{N\bh^2}}\left\langle
  \prod_{i=1}^{N}e^{i\bh\phi(\tau_i,x_i)}\prod_{i=1}^{N}
  e^{-i\bh\phi(\tau'_i,x'_i)}\right\rangle = \left[{\prod_{1\leq i < j\leq
        N} (z_i-z_j) (z'_i-z'_j) \over \prod_{1\leq i,j \leq
        N}(z_i-z'_j)}\right]^{\bh^2}= \left[\det M^{(N)}\right]^{\bh^2}.
\end{equation}
For the bosonic sequence it is clear from Eq.~(\ref{eq:2N-Dirac}) that
the $\bh^2$ power of the determinant can be reproduced by an
identification of the form
\begin{equation}
  \label{eq:multi-fermionization}
  e^{i\bh\phi(x)}\sim \prod_{i=1}^{{\bh^2}/2}\psi_i(x)
  \prod_{j={\bh^2}/2+1}^{\bh^2}\psi_j^{\dagger}(x),
\end{equation}
where the $\psi_i(x)$ are $\bh^2$ independent chiral Dirac fermions
with identical velocities.  This construction closely resembles Jain's
bulk parton construction for hierarchical fractional quantum Hall
states\cite{jain} and its edge analog\cite{wen}, although the
motivation here, a desire for a fermionic representation of the
tunneling operator, is very different.

In the \XSG Hamiltonian there is only a single chiral boson (with
central charge $c=1$), so we first have to add $(\bh^2-1)$ auxiliary
$c=1$ degrees of freedom (\eg, chiral bosons) in order to carry out
this procedure.  Of the points in the bosonic sequence, only the first
two ($\bh^2=2,4$) occur in the region where the cosine term is not
irrelevant [by Eq.~(\ref{eq:XSGflow})], and these are the ones we now
describe in detail.

\subsection{Relevant Tunneling, $\bh^2=2$}

As noted above, an additional degree of freedom is necessary for a
straightforward fermionization of the \XSG Hamiltonian, so we add to
the Hamiltonian an auxiliary free compact chiral boson
($\hat{\phi}$) with radius $\hat{R}$, topological charge $\hat{N}$,
and a velocity equal to the velocity of $\phi$.  The full
Hamiltonian is then:
\begin{eqnarray}
  {\hat {\cal H}}_{\chi SG}^{\bh^2=2}&\equiv&{\cal H}_{\chi
    SG}^{\bh^2=2}[\phi]+{\cal H}_F[\hat{\phi}] \nonumber \\
  &=&\ix\left[{{1\over 4\pi}v_{\rm n}:\!(\delx\phi)^2\!:+{1\over
        4\pi}v_{\rm n}:\!(\delx\hat{\phi})^2\!: +{\lambda\over 2\pi
        a}(e^{i\sqrt{2}\phi}+\,{\rm h.c.})}\right]. \label{eq:H_4pi.1}
\end{eqnarray}
We then perform a canonical transformation which mixes the field
appearing in the tunneling term with the auxiliary boson
\begin{equation}
  \label{eq:phi_to_theta}
  \pmatrix{\hat{\phi} \cr \phi \cr}={1\over \sqrt{2}}\pmatrix{1 & 1
    \cr 1 & -1 \cr} \pmatrix{\theta_1 \cr \theta_2},
\end{equation}
in terms of which we find
\begin{equation}
  \label{eq:H(theta)}
  {\hat {\cal H}}_{\chi SG}^{\bh^2=2}=\ix\left\{{{1\over
        4\pi}v_{\rm n}
      \,:\!\delx\theta_i \delx \theta_i\!:\,
      + {\lambda\over 2\pi
        a}\left[e^{i(\theta_1-\theta_2)}+\,{\rm h.c.}\right]}\right\}.
\end{equation}
The topological charge dependent part of the transformation in
Eq.~(\ref{eq:phi_to_theta}) is
\begin{equation}
  \label{eq:Nphi_to_Ntheta}
  {\hat R}{\hat
    N}= {1\over\sqrt{2}}
    (N_1^{\theta}R_1^{\theta}+N_2^{\theta}R_2^{\theta}),\qquad  
  {1\over \sqrt{2}}N_{\rm n}
    ={1\over\sqrt{2}}(N_1^{\theta}R_1^{\theta}-N_2^{\theta}R_2^{\theta}), 
\end{equation}
where $N_i^{\theta},R_i^{\theta}$ are the topological charge and
radius of the field $\theta_i(x)$, respectively.  We choose
$\hat{R}=1/\sqrt{2}$ so that the requirement $N_i^{\theta}\in \Zint$
is consistent with setting $R_i^{\theta}=1$.  These choices also
impose the gluing condition ${\hat N}+N_{\rm n}\in\Zint_{\rm even}$.  With
these values of the radii for the $\theta_i$ fields, we can define two 
chiral fermion operators
\begin{equation}
  \label{eq:newfermions}
  \psi_i(x)={1\over \sqrt{2\pi a}}e^{i\theta_i (x)}.
\end{equation}
These two flavors of fermions commute, but once again this can be
fixed without modifying what follows, and the details are given in
Appendix \ref{sec:Kleins}.  Using this definition in
Eq.~(\ref{eq:H(theta)}) we have a fermionized Hamiltonian
\begin{equation}
  \label{eq:H_4pifermi}
  {\hat {\cal H}}_{\chi SG}^{\bh^2=2}=\ix 
  \,:\!\left[{-iv_{\rm n}\psi_i^{\dagger}\delx\psi_i -\lambda
      (\psi_2^{\dagger}\psi_1+\,{\rm h.c.})}\right]\!:.
\end{equation}
As noted in Section \ref{sec:edgetheory}, the fermion fields introduced in
Eq.~(\ref{eq:newfermions}) obey antiperiodic boundary conditions and
therefore the allowed values of the momentum $k$ for the Fourier mode
operators $c_{ki}$ of $\psi_i(x)$ are $k\in (2\pi /L)(\Zint+1/2)$.
The fermionic normal ordering introduced in Eq.~(\ref{eq:H_4pifermi})
is defined with respect to the (non-degenerate) ground state of this
Hamiltonian at $\lambda=0$ which is annihilated by $c_{ki}$ for $k>0$
and $c^{\dagger}_{ki}$ for $k<0$.  By introducing 
$\psi_{\pm}\equiv (\psi_1\pm \psi_2)/\sqrt{2}$ the Hamiltonian is
diagonalized:
\begin{equation}
  \label{eq:H_4pi_final2}
  {\hat {\cal H}}_{\chi SG}^{\bh^2=2}=\ix
  \,:\!\left[{-iv_{\rm n}\left({\psi_+^{\dagger}\delx\psi_+ 
  + \psi_-^{\dagger}\delx\psi_-}\right) 
  -\lambda (\psi_+^{\dagger}\psi_+ +\psi_-^{\dagger}\psi_-)}\right]\!:.
\end{equation}

Note that in the case of the 110 state the charged boson
($\phi_{\rm c}$) has radius $R_{\rm c}=1/\sqrt{2}$ and hence the fermionization
can be accomplished without adding an auxiliary degree of freedom by
replacing $\hat{\phi}$ by $\phi_{\rm c}$ in Eq.~(\ref{eq:phi_to_theta}).
The result of this procedure is nothing more than the chiral fermion
description of the original Hamiltonian of two uncorrelated,
interacting, $\nu=1$ edges, given above in Eq.~(\ref{eq:H(110)}).
For the 110 state the advantage of using an auxiliary boson
rather than the charged mode in the fermionization is that
in the former case the theory is quadratic (\ref{eq:H_4pi_final2}),
whereas in the latter case it is not (\ref{eq:H(110)}).

Having solved the theory with the auxiliary boson, we now show at the
level of the partition function that projecting out the unphysical
degree of freedom from the spectrum of Eq.~(\ref{eq:H_4pi_final2})
reproduces the result from the direct solution in Section
\ref{sec:exactsolns}.  From Eq.~(\ref{eq:H_4pi_final2}) the partition
functions of the \XSG theory with the auxiliary boson is
\begin{equation}
  \label{eq:Z_XSG'_4pi}
  {\hat Z}^{\bh^2=2}_{\chi SG}(\beta)={1\over \varphi(q_{\rm n})^2}
  \sum_{N_{+},N_{-} \in \Zint} e^{-\beta \lambda 
  (N_{+}+N_{-})}q_{\rm n}^{{1\over 2}(N_{+}^2+N_{-}^2)}. 
\end{equation}
To write the partition function including the charged mode we must
again consider gluing conditions.  We must separate states according
to their $N_{\rm n}$-parity, which because of the condition ${\hat
  N}+N_{\rm n}\in\Zint_{\rm even}$, is equivalent in their ${\hat
  N}$-parity.  Since ${\hat N}$ is just the total fermion number
measured with respect to the $\lambda=0$ ground state, we see the
partition function of the full edge theory is:
\begin{equation}
  \label{eq:Z'_4pi}
  {\hat Z}^{\bh^2=2}(\beta)=\chi_{R_{\rm c}}^{(2(m+n),0)}(q_{\rm
  c}){\hat Z}_{\chi SG}^{\bh^2=2,\,{\rm even}} 
  +\chi_{R_{\rm c}}^{(2(m+n),m+n)}(q_{\rm c}){\hat Z}_{\chi
  SG}^{\bh^2=2,\,{\rm odd}}, 
\end{equation}
where ${\hat Z}_{\chi SG}^{\bh^2=2,\,{\rm even}}$ is given by
Eq.~(\ref{eq:Z_XSG'_4pi}) with the restriction $N_{+}+N_{-}\in\Zint_{\rm
  even}$ and similarly for ${\hat Z}_{\chi SG}^{\bh^2=2,\,{\rm odd}}$.  To project
out the radius $R=1/\sqrt{2}$ auxiliary boson we added (${\hat \phi}$)
in Eq.~(\ref{eq:H_4pi.1}) we must evaluate
\begin{equation}
  \label{eq:projZ_XSG}
  Z^{\bh^2=2}_{\chi SG}(\beta)={{\hat Z}_{\chi SG}^{\bh^2=2,\,{\rm even}} \over
  \chi_{1/\sqrt{2}}^{(2,0)}(q_{\rm n})} 
  + {{\hat Z}_{\chi SG}^{\bh^2=2,\,{\rm odd}} \over \chi_{1/\sqrt{2}}^{(2,1)}(q_{\rm n})}
\end{equation}
for the \XSG theory in isolation and
\begin{equation}
  \label{eq:projZ_4pi}
  Z^{\bh^2=2}(\beta)=\chi_{R_{\rm c}}^{(2(m+n),0)}(q_{\rm c}){{\hat
    Z}_{\chi SG}^{\bh^2=2,\,{\rm even}} 
    \over \chi_{1/\sqrt{2}}^{(2,0)}(q_{\rm n})} 
  +\chi_{R_{\rm c}}^{(2(m+n),m+n)}(q_{\rm c}) {{\hat Z}_{\chi
    SG}^{\bh^2=2,\,{\rm odd}} \over 
    \chi_{1/\sqrt{2}}^{(2,1)}(q_{\rm n})}, 
\end{equation}
for the full edge theory.  It is useful at this point to recall the
definitions of the Jacobi theta functions:
\begin{eqnarray}
  \Theta_2(\nu|\tau)&=& \sum_{r\in\Zint + {1\over 2}} e^{2\pi i \nu
    r}e^{i \pi \tau r^2}, \nonumber \\ 
  \Theta_3(\nu|\tau)&=& \sum_{r\in\Zint} e^{2\pi i \nu r}e^{i \pi \tau
    r^2}, \label{eq:Jacobi}  \\ 
  \Theta_4(\nu|\tau)&=& \sum_{r\in\Zint} (-1)^r e^{2\pi i \nu r}e^{i \pi
    \tau r^2}. \nonumber 
\end{eqnarray}
If we define ${\bar \nu}\equiv i\beta\lambda/2\pi$ and ${\bar \tau}
\equiv i\beta v_{\rm n}/L$, we see from the definition of the
chiral boson character (\ref{eq:CBchar}) and Eqns.~(\ref{eq:Jacobi}) that
\begin{equation}
  \label{eq:rewrite1}
  \chi_{1/\sqrt{2}}^{(2,0)}(q_{\rm n})={1\over \varphi(q_{\rm
  n})}\Theta_3(0|2\tb),\qquad  
  \chi_{1/\sqrt{2}}^{(2,1)}(q_{\rm n})={1\over \varphi(q_{\rm
  n})}\Theta_2(0|2\tb). 
\end{equation}
Next note that we can rewrite the sums appearing in ${\hat Z}_{\chi 
  SG}^{\bh^2=2,\,{\rm even}}$ and ${\hat Z}_{\chi SG}^{\bh^2=2,\,{\rm
  odd}}$ using 
\begin{eqnarray}
  \sum_{N_{+}+N_{-}\in\Zint_{{\rm even}}} &=&
  \sum_{N_{+},N_{-}\in\Zint}{1\over 2}\left[1+(-1)^{N_{+}+N_{-}}\right],
  \nonumber \\  
  \sum_{N_{+}+N_{-}\in\Zint_{{\rm odd}}}  &=&
  \sum_{N_{+},N_{-}\in\Zint}{1\over 2}\left[1-(-1)^{N_{+}+N_{-}}\right],
  \label{eq:rewrite2} 
\end{eqnarray}
and hence
\begin{eqnarray}
  {\hat Z}_{\chi SG}^{\bh^2=2,\,{\rm even}} &=& {1\over 2
  \varphi(q_{\rm n})^2} 
  \left\{{\left[{\Theta_3(\nb|\tb)}\right]^2
      +\left[{\Theta_4(\nb|\tb)}\right]^2}\right\}, 
  \nonumber \\ 
  {\hat Z}_{\chi SG}^{\bh^2=2,\,{\rm odd}}  &=& {1\over 2
  \varphi(q_{\rm n})^2} 
  \left\{{\left[{\Theta_3(\nb|\tb)}\right]^2
  -\left[{\Theta_4(\nb|\tb)}\right]^2}\right\}. \label{eq:rewrite3} 
\end{eqnarray}
Finally, using the standard doubling identities \cite{Bateman}
\begin{eqnarray}
  \Theta_3(2\nu|2\tau)&=&{ \left[{\Theta_3(\nu|\tau)}\right]^2
    +\left[{\Theta_4(\nu|\tau)}\right]^2 \over 2\Theta_3(0|2\tau)},
    \nonumber \\ 
    \Theta_2(2\nu|2\tau)&=&{ \left[{\Theta_3(\nu|\tau)}\right]^2 
      -\left[{\Theta_4(\nu|\tau)}\right]^2 \over 2\Theta_2(0|2\tau)},
    \label{eq:doubling} 
\end{eqnarray}
and Eqns.~(\ref{eq:rewrite1}) and (\ref{eq:rewrite3}) in
(\ref{eq:projZ_XSG}) and (\ref{eq:projZ_4pi}) we obtain
\begin{eqnarray}
  Z^{\bh^2=2}_{\chi SG}(\beta) &=&
  {1\over\varphi(q_{\rm n})}\left[{\Theta_3(2\nb|2\tb)
  +\Theta_2(2\nb|2\tb)}\right], 
  \nonumber \\  
  Z^{\bh^2=2}(\beta) &=& \chi_{R_{\rm c}}^{(2(m+n),0)}(q_{\rm
  c}){1\over\varphi(q_{\rm n})} 
  \Theta_3(2\nb|2\tb) 
  +\chi_{R_{\rm c}}^{(2(m+n),m+n)}(q_{\rm c}){1\over\varphi(q_{\rm
  n})}\Theta_2(2\nb|2\tb), 
  \label{eq:Z_4pi_final}
\end{eqnarray}
which by Eq.~(\ref{eq:Jacobi}) are in precise agreement with the
results of the exact solution in the previous section,
Eqns.~(\ref{eq:Z_XSG_4pi}) and (\ref{eq:Z_4pi}).  We have also checked
that correlation functions calculated with the fermionic 
Hamiltonian (\ref{eq:H_4pi_final2}) are identical to those found
in Eqns.~(\ref{eq:D_4pi}) and (\ref{eq:T_4pi}).

\subsection{Marginal Tunneling, $\bh^2=4$}

We next consider the generic multi-flavor fermionization of the \XSG
Hamiltonian at $\bh^2=4$.  We add to ${\cal H}_{\chi SG}^{\bh^2=4}$ the
Hamiltonian for three free chiral bosons ($\ph_i$) with radii ${\hat
  R}_i$ and topological charges ${\hat N}_i$:
\begin{eqnarray}
  {\hat {\cal H}}_{\chi SG}^{\bh^2=4}&\equiv&{\cal H}_{\chi
    SG}^{\bh^2=4}[\phi]+{\cal H}_F[\ph_1,\ph_2,\ph_3] \nonumber \\
    &=&\ix 
  \Biggl\{ {1\over 4\pi}
      \,:\!\biggl[v_{\rm n}(\delx
            \phi)^2 +\left({v_{\rm n}+{\lambda\over 2\pi}}\right)[(\delx
            \ph_1)^2+(\delx \ph_2)^2]
            \nonumber\\       & &  
           \qquad\qquad +\left({v_{\rm n}+{\lambda\over\pi}}\right)(\delx 
            \ph_3)^2\biggr]\!:
       {+{2\lambda\over (2\pi a)^2}\cos[2\phi(x)]}\Biggr\}.
  \label{eq:H_8pi}
\end{eqnarray}
The velocities of the auxiliary fields are completely at our
discretion, and the motivation for the specific choice made here will
become clear below.  We perform an orthogonal (canonical)
transformation which mixes the physical mode ($\phi$) with the
auxiliary bosons:
\begin{equation}
  \label{eq:phi_to_theta_2}
  \pmatrix{\phi \cr \hat{\phi}_1 \cr \hat{\phi}_2 \cr \hat{\phi}_3
    \cr}={1\over 2} \pmatrix{1 & -1 & 1 & -1 \cr \sqrt{2} & \sqrt{2} & 0 &
    0 \cr 0 & 0 & \sqrt{2} & \sqrt{2} \cr 1 & -1 & -1 & 1 \cr}
  \pmatrix{\theta_1 \cr \theta_2 \cr \theta_3 \cr \theta_4}.
\end{equation}
In terms of these new bosons our Hamiltonian is:
\begin{eqnarray}
  {\hat {\cal H}}_{\chi SG}^{\bh^2=4}&=&\ix \biggl\{ {1\over
        4\pi}\left({v_{\rm n}+{\lambda\over 2\pi}}\right):\!\delx
      \theta_i\,\delx\theta_i\!: +{\lambda\over (2\pi a)^2}
      \left[e^{i(\theta_1-\theta_2+\theta_3-\theta_4)}+\,{\rm
        h.c.}\right] \nonumber \\ & &+{\lambda\over
        2\pi}\,:\!
        (\delx\theta_1-\delx\theta_2)\,
        (\delx\theta_4-\delx\theta_3)\!:\biggr\}. 
  \label{eq:H_8pi(theta)}
\end{eqnarray}
The topological charge dependent part of the transformation in
Eq.~(\ref{eq:phi_to_theta_2}) is
\begin{eqnarray}
  {1\over 2}N_{\rm n}=&{1\over
    2}(R_1^{\theta}N_1^{\theta}-
    R_2^{\theta}N_2^{\theta}+R_3^{\theta}N_3^{\theta}-
    R_4^{\theta}N_4^{\theta}),\quad  
  {\hat R}_1{\hat N}_1=&{1\over
    \sqrt{2}}(R_1^{\theta}N_1^{\theta} +R_2^{\theta}N_2^{\theta}),
    \nonumber 
  \\ {\hat R}_3{\hat N}_3=&{1\over
    2}(R_1^{\theta}N_1^{\theta}
    -R_2^{\theta}N_2^{\theta}-R_3^{\theta}N_3^{\theta}+
    R_4^{\theta}N_4^{\theta}),\quad  
  {\hat R}_2{\hat N}_2=&{1\over
    \sqrt{2}}(R_3^{\theta}N_3^{\theta}+R_4^{\theta}N_4^{\theta}),
  \label{eq:Nphi_to_Ntheta_2}
\end{eqnarray}
where $N_i^{\theta},R_i^{\theta}$ are the topological charge and
radius of the field $\theta_i(x)$, respectively.  We again want to use
our freedom to choose the quantities ${\hat R}_i$ to ensure
$R_i^{\theta}=1$.  Clearly ${\hat R}_1={\hat R}_2=1/\sqrt{2},{\hat
R}_3=1/2$ accomplishes this.  Since the $\theta_i$ fields have unit
radius we can introduce a quartet of chiral fermion operators
\begin{equation}
  \label{eq:newfermions_8pi}
  \psi_i(x)={1\over \sqrt{2\pi a}}e^{i\theta_i(x)}.
\end{equation}
Once again these fermion operators are defined here without the Klein
factors necessary to ensure the proper anticommutation relations.
The fact that this can be remedied without modifying the form of the
Hamiltonian is demonstrated in Appendix \ref{sec:Kleins}.

In terms of these fermions Eq.~(\ref{eq:H_8pi(theta)}) reads:
\begin{eqnarray}
  {\hat {\cal H}}_{\chi SG}^{\bh^2=4}&=&\ix\,:\!\biggl\{{-iv'_{\rm
    n}\psi_i^{\dagger}\delx\psi_i 
    +{\lambda'\over 2} \Bigl[{ \psi_1 \psi_2^{\dagger} \psi_3
        \psi_4^{\dagger}+ \psi_4 \psi_3^{\dagger} \psi_2
        \psi_1^{\dagger}}}
    \nonumber\\ 
    & & {}+{{1\over 2}(\psi_1^{\dagger}\psi_1
    -\psi_2^{\dagger}\psi_2)\, 
    (\psi_4^{\dagger}\psi_4-\psi_3^{\dagger}\psi_3) 
       }\Bigr]\biggr\}\!:,
  \label{eq:H_fermi_8pi}
\end{eqnarray}
where $v'_{\rm n}\equiv v_{\rm n}+\lambda/2\pi$, and $\lambda' \equiv
2\lambda$.  The 
fermion normal ordering in Eq.~(\ref{eq:H_fermi_8pi}) is again defined
with respect to the $\lambda=0$ ground state which satisfies
$c_{ki}|\Omega\rangle=c_{-ki}^{\dagger}|\Omega\rangle=0$ for $k>0$ and
all $i$, where $k\in (2\pi/L)(\Zint +1/2)$ and $c_{ki}$ are the Fourier
components of $\psi_i(x)$.  At this value of $\bh^2$ the $\lambda=0$
ground state is the exact ground state of the Hamiltonian for
$\lambda\neq 0$ (in contrast to the $\bh^2=2$ case where the
$\lambda=0$ ground state is an exact eigenstate for $\lambda\neq 0$,
but not the ground state).
Assembling our four spinless fermions into a pair of (pseudo-)spin-1/2
fermions according to
\begin{equation}
  \label{eq:spin1/2}
  \Phi_1\equiv\pmatrix{\psi_1 \cr \psi_2 \cr},\qquad
  \Phi_2\equiv\pmatrix{\psi_4 \cr \psi_3 \cr},
\end{equation}
the Hamiltonian (\ref{eq:H_fermi_8pi}) can be written as a quadratic form
\begin{equation}
  \label{eq:H_8picurrent}
  {\hat {\cal H}}_{\chi SG}^{\bh^2=4}=\ix\,:\!\left[{ {\pi v'_{\rm n}\over
        2}(J_1^2+J_2^2)+ {2\pi v'_{\rm n}\over 3}(\bJ_1^2+\bJ_2^2)
      +\lambda'\,\bJ_1\!\cdot \bJ_2}\right]\!:
\end{equation}
in terms of the currents
\begin{equation}
  \label{eq:currents}
  J_i(x)\equiv\,
  :\!\Phi_{i\alpha}^{\dagger}(x)\,\Phi_{i\alpha}(x)\!:,\qquad 
  J_i^{a}(x)\equiv {1\over
    2}\,:\!\Phi_{i\alpha}^{\dagger}(x)\,\sigma^{a}_{\alpha\beta}\,
  \Phi_{i\beta}(x)\!:. 
\end{equation}
where $\sigma^{a}, a=x,y,z$, are the Pauli matrices.  The reason for
our choice of velocities in Eq.~(\ref{eq:H_8pi}) is now apparent:
along with the transformation in Eq.~(\ref{eq:phi_to_theta_2}) it
produces a combination of four-Fermi couplings in
Eq.~(\ref{eq:H_fermi_8pi}) that can be written as a scalar product of
currents.  This is not a unique construction, we can obtain the same
final form~(\ref{eq:H_8picurrent}) with a different choice of
velocities by modifying the transformation~(\ref{eq:phi_to_theta_2}).

Defining the Fourier components of the currents
\begin{equation}
  \label{eq:currentmodes}
  J_{i,n}\equiv\ix J_i(x)e^{-2\pi i n x/L},\quad \bJ_{i,n}\equiv\ix
  \bJ_i(x)e^{-2\pi i n x/L},\quad n\in\Zint,
\end{equation}
the algebra is
\begin{eqnarray}
  \left[{J_{i,n}, J_{j,m}}\right] & = &
  2n\delta_{ij}\delta_{n+m,0},\quad \left[{J_{i,n},J_{j,m}^{a}}\right]=0,
  \nonumber \\ \left[{J_{i,n}^{a}, J_{j,m}^{b}}\right] & = & {n\over
    2}\delta_{ij}\delta^{ab}\delta_{n+m,0} +i\delta_{ij}\epsilon^{abc}
  J_{i,n+m}^{c}. \label{eq:KMalgebras}
\end{eqnarray}
We therefore have two $\hat{u}(1)$ ($J_i$) and two $\widehat{su}(2)_1$ 
($\bJ_i$) Kac-Moody algebras.  From Eqns.~(\ref{eq:currents})
and (\ref{eq:currentmodes}) we find that the ground state satisfies
\begin{equation}
  \label{eq:J_on_g.s.}
  J_{i,n}|\Omega\rangle=J_{i,n}^{a}|\Omega\rangle=0,\,{\rm for}\quad
  i=1,2;\quad a=x,y,z;\quad n\geq 0,
\end{equation}
and therefore normal ordering for the current operator modes is
defined by moving any current with momentum index $n\leq 0$ to the
right past any current with index $m>0$ with which it has a
non-trivial commutator.

To solve Eq.~(\ref{eq:H_8picurrent}), the spin part of which is the
Sugawara form of two coupled $\widehat{su}(2)_1$ Wess-Zumino-Witten
models\cite{BYB}, we first note that we can define:
\begin{equation}
  \label{eq:Virasoros}
  {\cal L}_n^{(i)}\equiv{1\over 2}:\!J_{i,m}\,J_{i,n-m}\!:,\qquad
  L_n^{(i)}\equiv{1\over 3}:\!J_{i,m}^{a}\,J_{i,n-m}^{a}\!:.
\end{equation}
These four sets of operators obey four independent $c=1$ Virasoro
algebras.  The operators $\bJ_{+}\equiv \bJ_1+\bJ_2$ obey an
$\widehat{su}(2)_2$ diagonal subalgebra of $\widehat{su}(2)_1\oplus
\widehat{su}(2)_1$, and the bilinear operators formed from them
\begin{equation}
  \label{eq:su(2)2virasoro}
  L^{(+)}_n\equiv{1\over 4}:\!J_{+,m}^{a}\,J_{+,n-m}^{a}\!:,
\end{equation}
obey a $c=3/2$ Virasoro algebra.  From the GKO coset construction, the
operators $\Lt_n\equiv L_n^{(1)}+L_n^{(2)}-L_n^{(+)}$ form a $c=1/2$
Virasoro algebra which is independent of the one formed from the
$\bJ_{+,n}$ currents, \ie, $[\Lt_n,L_m^{(+)}]=0$.  Using these
observations and the trivial identity
\begin{equation}
  \label{eq:currentident}
  :\!\bJ_1(x)\cdot \bJ_2(x)\!:\,={1\over
    2}{:\!\left[{\bJ_+(x)}\right]^2\!:\over 4} -{3\over
    2}\left\{{:\!\left[{\bJ_1(x)}\right]^2\!:+:\!\left[{\bJ_2(x)}\right]^2\!:
      \over 3} -{:\!\left[{\bJ_+(x)}\right]^2\!:\over 4}\right\},
\end{equation}
the Hamiltonian becomes
\begin{equation}
  \label{eq:H_8pi_final}
  {\hat {\cal H}}_{\chi SG}^{\bh^2=4}={\pi v'_{\rm n}\over L}\left[{{\cal
        L}^{(1)}_{0}+{\cal L}^{(2)}_{0}}\right] +\left[{{2\pi v'_{\rm n}\over
        L}+{\lambda'\over 2 L}}\right]L^{(+)}_{0} +\left[{{2\pi v'_{\rm n}\over
        L}-{3\lambda'\over 2 L}}\right]\Lt_{0}.
\end{equation}
By transforming from $L^{(1)}_n, L^{(2)}_n$ to $L^{(+)}_n, \Lt_n$ we
have succeeded in diagonalizing the Hamiltonian.  From
Eqns.~(\ref{eq:J_on_g.s.}), (\ref{eq:Virasoros}),
(\ref{eq:su(2)2virasoro}), and the definition of $\Lt_n$ one finds that
the ground state $|\Omega\rangle$ is annihilated by ${\cal L}^{(i)}_n,
L^{(+)}_n, \Lt_n$ for $i=1,2$ and $n \ge 0$.  Excited states are
obtained by acting on $|\Omega\rangle$ with combinations of Virasoro generators
with $n<0$.

We now project out the auxiliary degrees of freedom we
added in Eq.~(\ref{eq:H_8pi}).  The Hilbert space of the \XSG
Hamiltonian is the subspace of the Hilbert space of ${\hat {\cal H}}_{\chi
SG}^{\bh^2=4}$ for which the auxiliary chiral bosons are in their
ground state.  We define a state $|\gamma\rangle$ as physical (\ie,
belonging to the Hilbert space of the \XSG Hamiltonian) if it
satisfies
\begin{equation}
  \label{eq:proj_ph}
  \ix 
  \,:\!(\delx \ph_i)^2\!:
  |\gamma\rangle =0,\;{\rm for}\quad i=1,2,3.
\end{equation}
Using Eqns.~(\ref{eq:phi_to_theta_2}), (\ref{eq:newfermions_8pi}),
(\ref{eq:spin1/2}), and (\ref{eq:currents}), these conditions can be
written in terms of currents as
\begin{eqnarray}
  \ix :\!\left[{J_i(x)}\right]^2\!:|\gamma\rangle&=&0\quad{\rm for}\quad
  i=1,2, \label{eq:proj_u(1)} \\ \ix
  :\!\left[{J_{+}^{z}(x)}\right]^2\!:|\gamma\rangle&=&0. \label{eq:proj_su(2)}
\end{eqnarray}
We note that these quadratic conditions can be recast as linear
constraints.  For example, using Eq.~(\ref{eq:currentmodes}) in
Eq.~(\ref{eq:proj_u(1)}) and acting on the left with $\langle\gamma|$
we find
\begin{equation}
  \label{eq:linear_proj}
  \|{J_{i,0}|\gamma\rangle}\|^2 + 2 \sum_{n=1}^{\infty}
  \|{J_{i,n}|\gamma\rangle}\|^2=0
\end{equation}
where we have used $\left({J_{i,n}}\right)^{\dagger}=J_{i,-n}$.  Since
the norm is positive semi-definite we see that
Eq.~(\ref{eq:proj_u(1)}) is equivalent to the set of linear conditions
$J_{i,n}|\gamma\rangle=0$ for $n\ge 0$.  The condition in
Eq.~(\ref{eq:proj_u(1)}) states that we should project out the
${\widehat u(1)}$ algebras, \ie, acting on any physical state with
${\cal L}^{(i)}_{-n}$ for $i=1,2$, and $n>0$, necessarily produces an
unphysical state.  From the Hamiltonian~(\ref{eq:H_8pi_final}), and
Eq.~(\ref{eq:proj_u(1)}) we have
\begin{equation}
  \label{eq:killu(1)'s}
  {\hat {\cal H}}_{\chi SG}^{\bh^2=4}|\gamma\rangle=\left({2\pi \over L}\right)
  \left[{\left({v'_{\rm n}+{\lambda'\over 4\pi}}\right)L^{(+)}_0
        +\left({v'_{\rm n} -{3\lambda'\over
  4\pi}}\right)}\Lt_0\right]|\gamma\rangle, 
\end{equation}
if $|\gamma\rangle$ is physical.  The second projection condition,
Eq.~(\ref{eq:proj_su(2)}), is slightly more subtle.  We are aided by
the observation that the operators $J_{+,n}^{z}$ obey a $\widehat
{u}(1)$ subalgebra of the $\widehat{su}(2)_2$ algebra generated by
$\bJ_{+,n}$.  Therefore
\begin{equation}
  \label{eq:ZVirasoro}
  L^{(z)}_n\equiv{1\over 2}:\!J_{+,m}^{z}\,J_{+,n-m}^{z}\!:,
\end{equation}
obey a $c=1$ Virasoro algebra and we can once again use the GKO
construction to define operators $\Lh_n\equiv L^{(+)}_n-L^{(z)}_n$
which are independent of the $L^{(z)}_n$ (\ie,
$[\Lh_n,L^{(z)}_m]=0$) and which obey a $c=1/2$ Virasoro algebra.
Using $L^{(+)}_0=\Lh_0+L^{(z)}_0$ and Eqns.~(\ref{eq:proj_su(2)}),
(\ref{eq:killu(1)'s}) and (\ref{eq:ZVirasoro}) we arrive at
\begin{equation}
  \label{eq:H_8pi_proj}
  {\hat {\cal H}}_{\chi SG}^{\bh^2=4}|\gamma\rangle={\cal H}_{\chi
    SG}^{\bh^2=4}|\gamma\rangle= \left({2\pi \over L}\right)
  \left[{\left({v'_{\rm n}+{\lambda'\over 4\pi}}\right)\Lh_0
        +\left({v'_{\rm n} -{3\lambda'\over
    4\pi}}\right)}\Lt_0\right]|\gamma\rangle, 
\end{equation}
if $|\gamma\rangle$ is a physical state.  One can readily show
$[\Lh_n,\Lt_m]=0$ and therefore from Eq.~(\ref{eq:H_8pi_proj}) we see
that the \XSG Hamiltonian can be written as the sum of two independent
$c=1/2$ pieces with velocities $(v'_{\rm n}+\lambda'/4\pi)=v_{\rm
  n}+\lambda/\pi=v_+$ and 
$(v'_{\rm n}-3\lambda'/ 4\pi)=v_{\rm n}-\lambda/\pi=v_-$, in precise
agreement with the 
expression~(\ref{eq:HXSG8pi,Maj}) obtained by the direct solution in
Section \ref{sec:exactsolns}, since a Majorana fermion has a central
charge of $1/2$.

\section{Fermionic Sequence}
\label{sec:fermiseq}

In Section \ref{sec:fermionizations} we discussed the generic
fermionization of the \XSG theory for points in the bosonic sequence
($\bh^2\in\Zint^{+}_{\rm even}$), which involves the introduction of
$(\bh^2-1)$ auxiliary chiral bosons.  This procedure can also be
applied to the fermionic sequence ($\bh^2\in\Zint^{+}_{\rm odd}$),
but the subtlety here is that the perturbation is antiperiodic,
from Eq.~(\ref{eq:period}) we have 
\beq
\cos\left({\bh\phi(x+L)}\right)=-\cos\left({\bh\phi(x)}\right),
\eeq
which is consistent with the fact that its scaling dimension is
half-integer and hence in the multi-flavor fermionization it is
represented as a product of an odd number of fermions.  One can modify
the definition of the theory at these points by adding a factor of
$e^{-i\pi x/L}$ to the tunneling amplitude $\lambda$ to make the
perturbation periodic.  For the case $\bh^2=1$ the chiral boson has
radius $R_{\rm n}=1$ and the theory can be fermionized without adding
auxiliary degrees of freedom.  With the modification to the tunneling
amplitude one finds
\beqarr
  {\cal H}_{\chi SG}^{\bh^2=1}&=& \ix\left[{{1\over 4\pi}v_{\rm n}
      \,:\!(\delx\phi)^2\!:\,
      +{\lambda\over \sqrt{2\pi a}}\left(e^{-i\pi x/L}e^{i\phi(x)}+\,{\rm
      h.c.}\right)}\right] \nonumber \\
   &=& \ix \,:\!\left[{-iv_{\rm n}\psi^{\dagger}\delx\psi 
       +\lambda \left({e^{-i\pi x/L}\psi(x)+e^{i\pi x/L}\psi^{\dagger}(x)
       }\right)}\right]\!: \nonumber \\ 
   &=& \left({2\pi v_{\rm n}\over
       L}\right)\sum_{r\in\Zint+1/2}r:c_r^{\dagger}c_r:
   +\lambda\sqrt{L} 
       (c_{1/2}+c_{1/2}^{\dagger}) \label{eq:HXSG2pi},
\eeqarr
where we have used $e^{i\phi(x)}/\sqrt{2\pi a}=\psi(x)=L^{-1/2}
\sum_r e^{2\pi i r x/L}c_r$.  A similar Hamiltonian, with the cosine
interaction at a point rather than along the line has appeared in
various physical contexts, see for example Ref.~\cite{matveev}.
Note that the perturbation changes only a single Fourier mode of the
chiral fermion.  Therefore in the limit $L\rightarrow
\infty$ the perturbation has no effect, which is to be expected since
a term in the Hamiltonian consisting of a product of an odd number
of fermions cannot produce any off-diagonal long range order.

\section{Semiclassical Considerations}
\label{sec:semiclass}

In the case of the standard sine-Gordon theory, the classical equation of 
motion has massive solitonic solutions.  A semiclassical analysis 
about these configurations, valid for small $\bh^2$, gives information 
about the spectrum of the quantum theory in the massive phase which 
exists for $\bh^2<4$ \cite{Rajaraman}.  In this section we demonstrate 
that the classical field theory of the \XSG model possesses analogous 
solitary wave (kink) solutions, and we address the question as to whether 
or not a similar semiclassical expansion is useful.

Consider the \XSG model (in the limit $L\rightarrow \infty$) as 
a classical field theory.  The real-time Lagrangian density is
\beq
\label{eq:ClassicalL}
{\cal L}_{\chi SG}=-{1\over 4\pi}\delx \phi (\delT\phi+ v_{\rm n}\delx \phi)
+\kappa(\cos(\bh \phi)-1),
\eeq
where the classical coupling constant $\kappa$ is related to the 
dimensionless coupling used in Section \ref{sec:RG} by 
$\kappa\equiv -\lh/(2\pi a)^2$.  We have included a constant term
in the potential energy so that the minima at $\phi=2\pi r/\bh$,
$r\in \Zint$ have zero energy.  The classical equation of motion is
\beq
\label{eq:EOM}
\delx(\delT +v_{\rm n} \delx)\phi - 2\pi \kappa\bh \sin(\bh \phi)=0.
\eeq
The conjugate momentum of the field $\phi(x)$ is $\Pi(x)=-\delx\phi(x)/4\pi$, 
and the Hamiltonian of the theory is
\beq
\label{eq:classH}
{\cal H}_{\chi SG}=\ixi \left[{ 4\pi v_{\rm n}\biglb(\Pi(x)\bigrb)^2
- \kappa (\cos(\bh \phi)-1) }\right].
\eeq
The equation of motion is reproduced in the Hamiltonian formalism by
defining the fundamental Poisson bracket of the chiral Bose field to
be\cite{FDMHunpub}
\beq
\label{eq:fundPB}
\{\delx \phi(x),\phi(x') \}_{PB}=2\pi \delta(x-x').
\eeq
This Poisson bracket is non-canonical, which is to be expected since
the system is constrained, \ie, $\Pi(x)$ can be expressed in terms of
$\phi(x)$.  This is in contrast to the standard SG theory, in which
the field and its conjugate momentum are independent dynamical
variables.  If we define the rescaled field ${\bar \phi}\equiv
\bh\phi$ to normalize the period of the potential energy term, we see 
from Eq.~(\ref{eq:fundPB}) that the Poisson bracket of $\delx{\bar \phi}(x)$ 
and ${\bar \phi}(x')$ is proportional to $\bh^2$.  Therefore we would expect
a semiclassical analysis to be relevant when $\bh^2$ is small, just 
as in the non-chiral SG model.  However, recall that the quantum \XSG 
theory at a finite length $L$ is only sensible for $\bh^2\in\Zint$, and 
hence in the region of $\bh^2$ where we expect the semiclassical expansion 
to be most valid, the quantum theory is problematic. 

If we consider only static field configurations, the equation of motion
is identical to the one obtained from the non-chiral SG theory.  Therefore,
Eq.~(\ref{eq:EOM}) has time-independent kink solutions which are
identical to those of the standard SG theory.  From these stationary
configurations we can find time-dependent solutions by applying a boost.
The result is:
\beq
\label{eq:soliton}
\phi_{\pm,v_s}(t,x)={4\over \bh}\tan^{-1}\left\{ {
\exp \left[{\pm \sqrt{2\pi \kappa \bh^2 \over v_{\rm n}-v_s}(x-v_s
    t)}\right]}\right\}, 
\eeq
where $v_s\in [-\infty,v_{\rm n}]$ is the velocity of the solitary wave.  
The topological charge is
\beq
\label{eq:classN}
N\equiv {\bh\over 2\pi} \ixi \delx\phi(x),
\eeq
and the solution $\phi_{\pm,v_s}$ has $N=\pm 1$.  These solitary waves
are not chiral, $v_s$ can be positive or negative, but there still is
left-right asymmetry.  The solitary waves that move to the right (the
chirality at $\kappa=0$) have a maximum velocity $v_{\rm n}$ and are
contracted relative to the $v_s=0$ solutions, while the solitary waves
moving to the left have no maximum velocity and are stretched relative
to the static solutions.  The existence of solitary wave solutions
with arbitrarily large velocities is in contrast to the standard SG
theory where the solitons have a maximum velocity because of the
Lorentz invariance of the theory.  Also, in the usual SG theory the
solitary wave solutions are solitons.  Whether or not this is true in
the \XSG theory is an open question.

Since ${\cal L}_{\chi SG}$ does not depend explicitly on $t$ or $x$ there
is a conserved energy, ${\cal E}$, and a conserved momentum, 
${\cal P}$, given by
\beqarr
{\cal E}[\phi]&=& \ixi \left[{ {v_{\rm n} \over 4\pi}(\delx \phi)^2
- \kappa (\cos(\bh \phi)-1) }\right] \label{eq:consE} \\
{\cal P}[\phi]&=& \ixi {1\over 4\pi}(\delx \phi)^2 \label{eq:consP}
\eeqarr
Using Eq.~(\ref{eq:soliton}) in Eqns.~(\ref{eq:consE}) and (\ref{eq:consP}), 
the energy and momentum of the solitary wave solutions are
\beqarr
E(v_s)& \equiv & {\cal E}[\phi_{\pm,v_s}]={v_{\rm n} \zeta\over
  \sqrt{v_{\rm n}-v_s}} 
+ \zeta \sqrt{v_{\rm n}-v_s} \label{eq:E(v)} \\
P(v_s)& \equiv & {\cal P}[\phi_{\pm,v_s}]={\zeta \over \sqrt{v_{\rm
      n}-v_s}} \label{eq:P(v)} 
\eeqarr
where $\zeta\equiv \sqrt{8\kappa/\pi \bh^2}$.  Note that $P(v_s)\geq
0$ even for solutions that propagate to the left ($v_s <0$).  The
dispersion relation is $E(P)=v_{\rm n} P + \zeta^2/P$, which has a minimum
at $P_0=\zeta/\sqrt{v_{\rm n}}$ for which $E(P_0)=2\sqrt{v_{\rm n}}\zeta$.

We have found that there are topologically non-trivial solitary wave
solutions of the classical equation of motion which exhibit a
dispersion relation with a finite energy gap of order $\sqrt{\kappa}$.
However, the quantum spectrum of the \XSG theory is gapless, at least
at the points $\bh^2=2$ and $\bh^2=4$.  It is unclear how a
semiclassical expansion about massive classical solutions could
possibly describe the massless quantum theory.  A second observation is
that while the solitary wave solutions in Eq.~(\ref{eq:soliton})
conserve the topological charge, the dynamics of the \XSG theory does
not.  From Eqns.~(\ref{eq:classH}), (\ref{eq:fundPB}), and
(\ref{eq:classN}) we see that
\beq
\label{eq:PBNH}
\{ N,{\cal H}_{\chi SG}\}_{PB}=\kappa\bh^2\ixi \sin \bh\phi(x).
\eeq
The functional on the r.h.s.\ of Eq.~(\ref{eq:PBNH}) is in general
non-zero, but it does vanish for $\phi_{\pm,v_s}$ since they are odd
functions of $(x-v_s t)$.  If at some instant an arbitrarily small
perturbation which destroys this symmetry is added to the solitary
wave solution, it follows from Eq.~(\ref{eq:PBNH}) that a naive time
evolution would result in a non-conserved winding number.  However,
from Eqns.~(\ref{eq:classN}) and (\ref{eq:consE}), we see that if the
topological charge of the system departs from an integer value, the
energy would be infinite.  Thus, if we impose boundary conditions on
the field and its spatial derivative such that the energy is finite,
specifically,
\beq
\label{eq:finiteE}
\lim_{x\rightarrow \pm\infty}\delx \phi(x)=0,\quad {\rm and}\quad 
\lim_{x\rightarrow \pm\infty}\phi(x)\in \left({2\pi\over\bh}\right)\Zint,
\eeq
then the winding number will be restricted to the integers, and hence
it will be conserved if we assume continuity of $\phi(t,x)$.  In the
quantum \XSG theory the topological charge is {\em not\/} conserved, as
can be seen from the non-vanishing of the commutator corresponding to
the Poisson bracket in Eq.~(\ref{eq:PBNH}).  We therefore believe a
semiclassical expansion about the solitary wave solutions is unlikely
to be useful for understanding the spectrum of the quantum theory.

\section{Summary}
\label{sec:disc}

We have considered the edge theory of double-layer quantum Hall
systems with a single mode per edge, upon inclusion of interlayer
single-electron tunneling at the edge.  The theory can be separated
into a free chiral boson Hamiltonian for the charged mode and a chiral
sine-Gordon Hamiltonian for the neutral mode.  We have argued that the
RG flow of the \XSG theory differs significantly from that of the
standard, non-chiral sine-Gordon theory, in particular $\bh$ is not
renormalized and there is a line of fixed points at $\bh^2=4$.  In
addition, the \XSG theory likely only makes sense for integer values
of $\bh^2$.

These values in turn break up into an apparently unphysical fermionic
sequence, and a bosonic sequence that is realized in quantum Hall
systems.  In the bosonic sequence there are two points at which
tunneling is not irrelevant, $\bh^2=2$ and $\bh^2=4$, and each
corresponds to an infinite set of double-layer quantum Hall edge
theories.  For these two cases we have obtained exact solutions for
the partition function and some correlation functions.  At $\bh^2=2$
we find that tunneling produces spatial oscillations in the
correlation functions while at $\bh^2=4$ tunneling splits the
velocities of the two Majorana components of the chiral Dirac fermion
corresponding to the neutral mode.

The exact solution of the $\bh^2=4$ case involves an unfamiliar
Bose-Fermi identity, $\psi\delx\psi\sim e^{2i\phi}$, which is
put on firm ground in the appendices.  This is the first in a
sequence of identities of the form
\begin{equation}
  \label{eq:moreBFident}
  e^{ik\phi(x)}\sim \prod_{m=0}^{k-1} \partial_x^{m}\psi(x),\qquad
  k\in\Zint^{+}.
\end{equation}

Extensions of this work to include the effects of disorder, asymmetric
velocities $v\neq v'$, and the case of more than two layers are
currently being pursued\cite{nps-wip} and we expect to discuss these,
along with a proposal to detect the trifurcation of the electron at
the 331 edge, in a future publication.

\section{Acknowledgments}
\label{sec:ack}
We would like to thank E. Fradkin, I. Gruzberg, I. Klebanov, R. Konik,
D.-H. Lee, M. Stone, H. Verlinde, and X.-G. Wen for useful
discussions.  We thank H. Saleur for bringing Ref.~\cite{Dijkgraaf} to
our attention.  We would like to acknowledge support by a NSF Graduate
Research Fellowship (JDN), DOE Grant DE-FG02-90ER40542 (LPP) as well
as NSF grant No. DMR-96-32690, US-Israel BSF grant No. 9600294,
and fellowships from the A. P. Sloan
Foundation and the David and Lucille Packard Foundation (SLS).

{\em Note added:} After completion of this work, F.~D.~M. Haldane
pointed out to us that identities similar to Eq.~(\ref{eq:fermie2phi}) 
are implicit in work involving Umklapp scattering for spinless electrons
in one dimension\cite{FDMH80,black&emery}. As we have been unable to
find an explicit proof in the literature we have chosen to retain our
proofs in the interests of completeness.

\appendix

\section{Klein Factors}
\label{sec:Kleins}

In this appendix we demonstrate that it is possible to modify our
definitions of fermionic vertex operators in a way that gives the
proper anticommutation relations between different flavors without
altering the form of the Hamiltonian.  The first case to consider is
the physical electron operators defined in Eq.~(\ref{eq:elec_ops}),
which commute for electrons in different layers.  We modify this
definition to
\begin{equation}
  \label{eq:elec_opsw/Kleins}
  \Psi_i(x)={1\over L^{m/2}} e^{i\pi S_{ij}{\cal N}_j}e^{i u_i(x)},
\end{equation}
where $S$ is a matrix used to enforce the correct anticommutation
relations.  We now have, using Eqns.~(\ref{eq:varphi_CR}) and
(\ref{eq:N,varphi_CR}),
\begin{equation}
  \label{eq:electronCR}
  \Psi_i(x)\Psi_j(x')=\Psi_j(x')\Psi_i(x) e^{i\pi [{\rm sgn}(x'-x)
    K_{ij}-S_{ik}K_{kj}+S_{jk}K_{ki}]}\qquad \hbox{for}\qquad x\neq x'.
\end{equation}
If $K_{12}=n$ is odd we can set $S=0$ and have the correct
anticommutation relations between the electron operators in different
layers.  If $n$ is even then we require
\begin{equation}
  \label{eq:requireACR}
  n(S_{11}-S_{22})+m'S_{12}-mS_{21}\in \Zint_{\rm odd}.
\end{equation}
From the dimension of the operator in
Eq.~(\ref{eq:elec_opsw/Kleins}) it is clear that for $m\neq 1$ it is
not possible for $\Psi_i(x)$ and $\Psi^{\dagger}_i(x')$ to have a
canonical anticommutator; a characteristic consequence of a
low-energy projection.  It can be shown that
$\{:\!\Psi_i(x)\!:,\,:\!\Psi^{\dagger}_j(x')\!:\} \propto
\delta_{ij}\delta^{(m-1)}(x-x')$ in the limit $a\rightarrow 0,
L\rightarrow \infty$.

Using Eq.~(\ref{eq:elec_opsw/Kleins}) in the definition of the tunneling
Hamiltonian~(\ref{eq:H_1}) gives:
\begin{equation}
  \label{eq:H_1 w/Kleins}
  {\cal H}_1={\lambda_0\over L^m} \ix (:\!e^{i\pi
    \left({S_{2j}(K_{j2}-K_{j1})-(S_{1j}-S_{2j}){\cal N}_j}\right)}
  e^{i u_1(x)}e^{-i u_2(x)}\!: +\, {\rm h.c.}).
\end{equation}
The topological-charge dependent coefficient is just $\pm 1$ provided:
\begin{equation}
  \label{eq:noKleinsinH_1}
  S_{12}-S_{22},\quad S_{11}-S_{21} \in \Zint_{\rm even}.
\end{equation}
For any even $n$ and odd $m,m'$ there are infinitely many choices for
$S$ that simultaneously satisfy Eq.~(\ref{eq:requireACR}) and
Eq.~(\ref{eq:noKleinsinH_1}).  We make the particularly simple choice
\begin{equation}
  \label{eq:S}
  S=\pmatrix{ 1 & 0 \cr 1 & 0 \cr}\qquad \hbox{for}\qquad n \in
  \Zint_{\rm even},
\end{equation}
and absorb the overall sign into the definition of the tunneling
amplitude $\lambda\rightarrow (-1)^{n+1}\lambda$, making
Eq.~(\ref{eq:H_1 w/Kleins}) identical to Eq.~(\ref{eq:H_1}).

The second place where Klein factors are needed is in the
fermionizations of the \XSG Hamiltonian in Section
\ref{sec:fermionizations}.  For the case $\bh^2=2$,
Eq.~(\ref{eq:newfermions}) should be modified to read
\begin{equation}
  \label{eq:newfermionsw/Kleins}
  \psi_i(x)={1\over \sqrt{2\pi a}}e^{i\pi
    S_{ij}N_j^{\theta}}e^{i\theta_i (x)}.
\end{equation}
The matrix $S$ is again given by Eq.~(\ref{eq:S}) and ensures the
proper anticommutation relations between the different fermion species
without modifying Eq.~(\ref{eq:H_4pifermi}), just as in
Eq.~(\ref{eq:H_1 w/Kleins}).  For the multi-flavor fermionization at
$\bh^2=4$, Eq.~(\ref{eq:newfermions_8pi}) should be replaced by
\begin{equation}
  \label{eq:newfermions_8piwKleins}
  \psi_i(x)={1\over \sqrt{2\pi a}}e^{i \pi
    P_{ij}N^{\theta}_j}e^{i\theta_i(x)},
\end{equation}
where the matrix $P$ is chosen so that $\psi_i$ and $\psi_j$
anticommute for $i\neq j$.  One can readily show:
\begin{equation}
  \label{eq:newfermionCR}
  \psi_i(x)\psi_j(x')=\psi_j(x')\psi_i(x)e^{i\pi(P_{ji}-P_{ij} +{\rm
  sgn}(x-x')\delta_{ij})} 
  \quad\hbox{for}\quad x\neq x',
\end{equation}
and therefore we demand
\begin{equation}
  \label{eq:restrictP}
  P_{ij}-P_{ji}\in \Zint_{\rm odd}\quad\hbox{for}\quad i \neq j.
\end{equation}
Using Eq.~(\ref{eq:newfermions_8piwKleins}) in
Eq.~(\ref{eq:H_8pi(theta)}) we find the fermionized Hamiltonian
including Klein factors is
\begin{eqnarray}
  {\hat {\cal H}}_{\chi SG}^{\bh^2=4}&=&\ix:\!\biggl[{-iv'_{\rm
    n}\psi_i^{\dagger}\delx\psi_i + 
    \lambda(e^{i\pi p}e^{i\pi O_i P_{ij}N_{j}}\psi_1 \psi_2^{\dagger}
    \psi_3 \psi_4^{\dagger} + {\rm
      h.c.})}\nonumber \\ & &
 \qquad\qquad\qquad
    +{\lambda\over2}(\psi_1^{\dagger}\psi_1-\psi_2^{\dagger}\psi_2)   
  (\psi_4^{\dagger}\psi_4-\psi_3^{\dagger}\psi_3) \biggr]\!:,
  \label{eq:H_fermi_8piwKleins}
\end{eqnarray}
where $O_i=(-1)^i$ and
$p=P_{21}-P_{22}+P_{32}-P_{31}+P_{41}-P_{42}+P_{43}-P_{44}$.  From
Eqns.~(\ref{eq:restrictP}) and (\ref{eq:H_fermi_8piwKleins}) we see
that one choice of $P$ that gives the correct anticommutation
relations and which trivializes the explicit topological charge
dependence of the Hamiltonian is
\begin{equation}
  \label{eq:P}
  P=\pmatrix{0 & 1 & 1 & 1 \cr 0 & 1 & 1 & 1 \cr 0 & 0 & 0 & 1 \cr 0 & 0
    & 0 & 1 \cr}.
\end{equation}
With this choice of $P$, we find Eq.~(\ref{eq:H_fermi_8piwKleins}) is
identical to Eq.~(\ref{eq:H_fermi_8pi}).

\section{Equivalence of Correlation Functions}
\label{sec:corr}

In this appendix we prove the validity of the identity
\begin{equation}
  \label{eq:bosonID}
  \psi(x)\delx\psi(x)={i\over 2\pi a^2}e^{2i\phi(x)},
\end{equation}
by comparing the $2N$-point correlation functions of these operators
and their Hermitian conjugates.  We work at zero temperature for
simplicity.  Consider first the correlation function in the
free chiral boson theory:
\begin{eqnarray}
  {\cal B}_N(z_1,\ldots ,z_N;z'_1,\ldots ,z'_N)& \equiv & \left\langle{
  \prod_{i=1}^{N} \left({i\over{2\pi a^2}}\right) e^{2i\phi(\tau_i,x_i)}
  \prod_{i=1}^{N} \left({-i\over{2\pi a^2}}\right)
  e^{-2i\phi(\tau'_i,x'_i)}}\right\rangle \nonumber \\ & = & {1\over
    (2\pi)^{2N}} { \prod_{1\leq i < j\leq N} (z_i-z_j)^4 (z'_i-z'_j)^4
    \over \prod_{1\leq i,j \leq N}(z_i-z'_j)^4}. \label{eq:B}
\end{eqnarray}
From Eq.~(\ref{eq:bosonID}), the corresponding correlation function in 
the free chiral fermion theory is:
\begin{equation}
  \label{eq:F}
  {\cal F}_N(z_1,\ldots ,z_N;z'_1,\ldots ,z'_N)\equiv \left\langle{
  \prod_{i=1}^{N}\psi(\tau_i,x_i)\partial_{x_i}\psi(\tau_i,x_i)\prod_{i=1}^{N}
  \left({-\psi^{\dagger}(\tau'_i,x'_i)\partial_{x'_i}
      \psi^{\dagger}(\tau'_i,x'_i)}\right)}\right\rangle.
\end{equation}
We adopt the complex notation
$\psi(\tau,x)\equiv\psi(z),\delx\psi(\tau,x)\equiv
-i\partial_z\psi(z)$.  To evaluate this $2N$-point function we write
it as the limit of a $4N$-point function and pull the derivatives
outside the correlator:
\begin{eqnarray}
  {\cal F}_N(\{z_i\},\{z'_i\}) & = & \lim_{w_1,w_2\rightarrow z_1}\quad
  \lim_{w'_1,w'_2\rightarrow z'_1} \ldots
  \lim_{w_{2N-1},w_{2N}\rightarrow z_N}\quad
  \lim_{w'_{2N-1},w'_{2N}\rightarrow z'_N} \nonumber \\ & &
  \left[{\prod_{j=1}^{N}(\partial_{w_{2j}}\partial_{w'_{2j}})
        \left\langle{\prod_{i=1}^{2N}
  \psi(w_i)\prod_{i=1}^{2N}\psi^{\dagger}(w'_i)}\right\rangle}\right].  
  \label{eq:splitting}
\end{eqnarray}
From Eqns.~(\ref{eq:cauchy}) and (\ref{eq:2N-Dirac}) we find that the
$4N$-point correlation 
function can be written:
\begin{equation}
  \label{eq:DiracDet}
  \left\langle{\prod_{i=1}^{2N}
  \psi(w_i)\prod_{i=1}^{2N}\psi^{\dagger}(w'_i)}\right\rangle={1\over 
  (2\pi)^{2N}} 
  {\prod_{1\leq i < j\leq 2N} (w_i-w_j)(w'_i-w'_j) \over \prod_{1\leq
  i,j \leq 2N}(w_i-w'_j)}. 
\end{equation}

We first organize the products appearing in Eq.~(\ref{eq:DiracDet}) by
separating coordinates, $w_i$, with even and odd indices.  The
denominator can be expressed as
\begin{equation}
  \label{eq:denom}
  \prod_{i=1}^{2N}\prod_{j=1}^{2N}
  (w_i-w'_j)=\prod_{i=1}^{N}\prod_{j=1}^{N}\left[{ 
      (w_{2i-1}-w'_{2j-1})(w_{2i}-w'_{2j-1})(w_{2i-1}-w'_{2j})(w_{2i}-w'_{2j})}
  \right],  
\end{equation}
and the numerator can be rewritten using
\begin{eqnarray}
  \prod_{i=1}^{2N-1}
  \prod_{j=i+1}^{2N}(w_i-w_j)&=&\prod_{i=1}^{N-1}\prod_{j=i+1}^{N}\biggl[
    (w_{2i-1}-w_{2j-1})(w_{2i}-w_{2j-1})\nonumber\\
    & & \times
    (w_{2i-1}-w_{2j})(w_{2i}-w_{2j})\biggr]
  \prod_{k=1}^{N}(w_{2k-1}-w_{2k}),
\label{eq:numer}
\end{eqnarray}
and an identical expression with $w_i$ replaced by $w'_i$.  

When the $N^2$ derivatives in Eq.~(\ref{eq:splitting}) act on the
expression on the r.h.s.\ of Eq.~(\ref{eq:DiracDet}), a large number of
terms are generated.  However, from Eq.~(\ref{eq:numer}) we see that
the numerator in Eq.~(\ref{eq:DiracDet}) contains a factor
$(w_{2k-1}-w_{2k})$ for each $k=1,\ldots N$.  In the limit
$w_{2k-1},w_{2k}\rightarrow z_k$, this factor vanishes, and hence the
only term that will be non-zero in the multiple limit given in
Eq.~(\ref{eq:splitting}) is the term in which the operator
$\partial_{w_{2k}}$ acts on $(w_{2k-1}-w_{2k})$ for all $k=1,\ldots
N$.  An analogous argument holds for the primed coordinates and
therefore we find using Eqns.~(\ref{eq:B}), and
(\ref{eq:splitting})-(\ref{eq:numer}): 
\begin{eqnarray}
\lefteqn{{\cal F}_N(\{z_i\},\{z'_i\})  =  \lim_{w_1,w_2\rightarrow z_1}\quad
  \lim_{w'_1,w'_2\rightarrow z'_1} \ldots
  \lim_{w_{2N-1},w_{2N}\rightarrow z_N}\quad
  \lim_{w'_{2N-1},w'_{2N}\rightarrow z'_N} } & & \nonumber \\
 \quad & &\times  {1\over (2\pi)^{2N}}{ \prod_{1 \leq i < j \leq N}\left[{
  (w_{2i-1}-w_{2j-1})
  (w_{2i}-w_{2j-1})(w_{2i-1}-w_{2j})(w_{2i}-w_{2j})}\right]\times  
  [w_i\rightarrow w'_i] \over
  \prod_{1 \leq i,j \leq N}(w_{2i-1}-w'_{2j-1})
  (w_{2i}-w'_{2j-1})(w_{2i-1}-w'_{2j})(w_{2i}-w'_{2j})} \nonumber \\  
  & = & {1\over (2\pi)^{2N}}{\prod_{1\leq i < j \leq N}(z_i-z_j)^4
    (z'_i-z'_j)^4 \over  
  \prod_{1\leq i, j \leq N} (z_i-z'_j)^4 }= {\cal
  B}_N(\{z_i\},\{z'_i\}), \label{eq:evenodd} 
\end{eqnarray}
completing the proof.

\section{Equivalence of Commutation Relations}
\label{sec:comm}

Here we calculate some additional commutation relations in support
of the identification~(\ref{eq:fermie2phi}), which can be written in
terms of a normal-ordered exponential [using
Eq.~(\ref{eq:normal-ordering})] as 
\begin{equation}
  \label{eq:eight-pi-no}
  {2\pi i\over L^2}\,:\!e^{2i\phi}\!:\,=\psi\,\partial_x\psi. 
\end{equation}
Consider the operators 
\begin{eqnarray}
  \label{eq:Am-defined}
  \Ab{p-1}(x)&\equiv&{1\over p\,k_1^{p}}
  \,:\!\left(\partial_x\phi\right)^p\!:\, =\sum_n
  \Ab{p-1}_n\,e^{ik_nx}, \\ 
  \label{eq:Cm-defined}
  \Cb\pm(x)&\equiv&{1\over2}\, :\!\left(e^{2i\,\phi}\pm
  e^{-2i\,\phi}\right)\!:\,  = \sum_n
  \Cb\pm_n\,e^{ik_nx},
\end{eqnarray}
where $k_1\equiv 2\pi/L$ and the rightmost sides of the expressions
define the normalization of the associated harmonics.  One can check
that
\begin{eqnarray}   
  \label{eq:comm-Cb-Ab0}
  \left[\Cb\pm_n,\, \Ab{0}_m\right] &=&-2\,\Cb\mp_{n+m},\\
  \label{eq:comm-Cb-Ab1}
  \left[\Cb\pm_n,\, \Ab{1}_m\right] &=&(n-m)\,\Cb{\pm}_{n+m},\\
  \label{eq:comm-Cbp-Cbp}
  \left[\Cb\pm_n,\, \Cb{\pm}_m\right]
  &=&\left(n-m\right)\,\Ab{1}_{n+m}+{1\over12}n\,(n^2-1)\,\delta_{n+m},\\
  \label{eq:comm-Cbp-Cbm}
  \left[\Cb+_n,\, \Cb-_m\right]
  &=&{2}\Ab2_{n+m}+ 
  {n^2-n\,m+m^2-1\over6}\Ab0_{n+m}.
\end{eqnarray}
For example, to derive Eq.~(\ref{eq:comm-Cbp-Cbp}), we write each
operator as a Fourier transformation,
$$ %
\Cb\pm_n=L^{-1}\int_{-L/2}^{L/2} {dx}\,e^{-i k_n\,x}\,\Cb\pm(x),
$$ %
and evaluate the products of normal ordered vertex operators with the
help of decomposition~(\ref{eq:sepvertex}) explicitly at a given
cutoff $a>0$,
\begin{equation}
  :\!e^{i\alpha_1\phi(x)} \!:\,\,:\! e^{ i\alpha_2\phi(y)}\!:\,
  =\,:\!e^{i\alpha_1\phi(x)+ i\alpha_2\phi(y)}\!: \,
  {\displaystyle e^{ -\pi\,i\,\alpha_1\alpha_2 \,(y-x)/L}\over
    \displaystyle
    \left[1-e^{2\pi\,i (y-x+i
          a)/L}\right]^{-\alpha_1\alpha_2}}. 
  \label{eq:svd-beta-two}
\end{equation}
In the limit $a\to0$, the commutator of these two vertex operators
vanishes at $x\neq y$ for even integer values of the product
$\alpha_1\alpha_2$, but it has a pole of order
$|\alpha_1\alpha_2|$ at $x=y$ for negative values of the
product, $\alpha_1\,\alpha_2<0$.  A direct calculation
with the help of the substitution $z=\exp(2\pi\,i\, x/L)$ shows that the
only contribution to the commutator~(\ref{eq:comm-Cbp-Cbp}) is given
by the corresponding residue,
\begin{eqnarray}
  \label{eq:comm-general-two-vertex}
  \left[\Cb+_n,\,
  \Cb{+}_m\right]&=&{1\over4}\sum_{\alpha=\pm2}\,\, 
  \oint\limits_{|z|=1} {dz\,z^{\alpha^2/2-n -1}\over (2\pi i)\,
    (\alpha^2-1)!}  
  \,{\partial^{(\alpha^2-1)}\over \partial 
    w^{(\alpha^2-1)}}\left[w^{\alpha^2/2-m-1}\,
      :\!e^{i\alpha\,(\varphi_z-\varphi_w)}\!:\right]\biggr|_{w=z}
  \\
  &=&{1\over24}\sum_{\alpha=\pm2}\oint{dz\,z^{1-n}\over 2\pi i}\,
  {\partial^3\over \partial w^3}\left[w^{1-m} 
      \,:\!e^{i\alpha\,(\varphi_z-\varphi_w)}\!:\right]\biggr|_{w=z} 
  \nonumber\\
  & &\hskip-0.9in=  \oint\limits_{|z|=1} {dz\,z^{-n-m-1}\over 2\pi i}
  \left[{m\,(1-m^2)\over12}
  +(m-1)\,z^{2}\,:\!(\partial_z\varphi_z)^2\!:-\,
   { z^{3}\over2}\partial_z:\!(\partial_z\varphi_z)^2\!:\,\right],
  \label{eq:residue-contribution}
\end{eqnarray}
and a simple integration by parts gives the
result~(\ref{eq:comm-Cbp-Cbp}).  In
Eq.~(\ref{eq:residue-contribution}) we used [{\em cf}.
  Eq.~(\ref{eq:phi_modes})]
$$
\varphi_z= -NR\,i\ln z-\chi/R+\sum_{r=1}^{\infty}{1\over\sqrt r}\,e^{-k_r
  a/2}\left(z^{-r}\,b_r^\dagger +z^r\,b_r\right).
$$
Readers familiar with conformal field theory techniques will notice that
Eq.~(\ref{eq:comm-general-two-vertex}) could be obtained immediately
within the formalism of canonical quantization on a
cylinder\cite{BYB}. We also notice that Eq.~(\ref{eq:comm-Cb-Ab1})
follows from a more general relationship for the vertex op\-e\-ra\-tor
${\cal V}^{\alpha}(x)\equiv \,:\!e^{i\alpha\phi}\!:$,
\begin{equation}
  \label{eq:comm-Verb-Ab1}
    \left[{\cal V}^{\alpha}_n,\, \Ab{1}_m\right] =
    \left(n+m-m\,{\alpha^2\over2}\right)\,
    {\cal V}^{\alpha}_{n+m},
\end{equation}
which can be derived in a variety of standard ways.  

On the other hand, using Eq.~(\ref{eq:eight-pi-no}) and the canonical
bosonization prescription, we can define the fermionic counterparts of
the same operators,
\begin{eqnarray*}
  \Af{p-1}(x)&\equiv& {L^p\over
    p}\,:\!{\left(\,:\!\psi^\dagger\psi\!:\,\right)^p}\!: ,\\
  \Cf{\pm}(x)&\equiv&{L^2\over4\pi i}\left(
    \psi\,\partial_x\psi\mp\partial_x\psi^\dagger\,\psi^\dagger
  \right) ={1\over2}\sum_r e^{i\,k_r x} \sum_s s \left( c_{r-s} c_s\pm
    c_s^\dagger c_{s-r}^\dagger\right),
\end{eqnarray*}
where the canonical Fermi operators $c_r$,
$\{c_r,\,c^\dagger_s\}=\delta_{rs}$ are defined by the expansion of
the antiperiodic field
$$
\psi(x)=L^{-1/2}\sum_{r\in\Zint+{1\over 2}} c_r\,e^{2\pi i r x/L}.
$$
Comparing their commutation relations, it is a straightforward
(but lengthy) exercise to express the operators
$$
\Af0_n=\Rf{0,0}_n,\quad
\Af1_n={1\over2}\left(\Rf{1,0}_n+\Rf{0,1}_n\right),\quad
\Af2_n\equiv \Rf{1,1}_n+{2n^2+1\over6}\,\Rf{0,0}_n,\ldots
$$
in terms of operators
$$
\Rf{a,b}_{n}=\lim_{K\to\infty}\sum_{r=-K}^{K+n}(r-n)^a\,
r^b\,:\!c_{r-n}^\dagger c_r\!:,
$$
bilinear in fermions.  The commutation relations then read
\begin{eqnarray*}
  {\left[\Cf{\pm}_n,\,\Af0_m\right]}&=&-2 \Cf{\mp}_{n+m},
  \\
  {\left[\Cf{\pm}_n,\,\Af1_m\right]}&=&(n-m)\,\Cf{\pm}_{n+m} ,
  \\
  {\left[\Cf{\pm}_n,\,\Cf{\pm}_m\right]}&=&
  (n-m)\,\Af1_{n+m}+{1\over12}n\,(n^2-1)\,\delta_{n+m},
  \\
  {\left[\Cf{+}_n,\,\Cf{-}_m\right]}&=&
  2\Rf{1,1}_{n+m}+{(n+m)^2-n\,m\over2}\Rf{0,0}_{n+m}
  \\
  &=& 2\Af2_{n+m} +{n^2-n\,m+m^2-1\over6}\Af0_{n+m}; 
\end{eqnarray*}
they are in precise correspondence with
Eqns.~(\ref{eq:comm-Cb-Ab0})--(\ref{eq:comm-Cbp-Cbm}).

\section{Alternative Solution of the $\chi$SG Theory at $\bh^2=4$.}
\label{sec:direct}

With the help of the commutation relations derived in
Appendix~\ref{sec:comm}, we can diagonalize the \XSG Hamiltonian at
$\bh^2=4$, Eq.~(\ref{eq:newHXSG8pi}), directly within the operator
formalism.  Indeed, in terms of normal-ordered quantities the
Hamiltonian reads
\begin{eqnarray}
  {\cal H}_{\chi SG}^{\bh^2=4}
  &=&\ix \left[{v_{\rm n}\over 4\pi}:\!(\delx \phi)^2\!: 
      +{2\lambda\over L^2}:\!\cos(2\phi)\!:\right]
  \nonumber\\
  & =&{2\pi\over L^2}\ix \left[ {v_{\rm n}}\,A^{(1)}(x)
      +{\lambda\over\pi}\,C^{(+)}(x)\right] \\
  &=&  {2\pi\over L}\left[v_{\rm n} A^{(1)}_0+{\lambda \over \pi}
  C^{(+)}_0\right],  
  \label{eq:csg-hamilt-two}
\end{eqnarray}
where we used the normal ordering formula~(\ref{eq:normal-ordering})
and the definitions~(\ref{eq:Am-defined}), (\ref{eq:Cm-defined}); the
last line is written as a sum of two commuting operators.  The commutation
relations~(\ref{eq:comm-Cb-Ab0})--(\ref{eq:comm-Cbp-Cbm}) imply
that the operators
$$
L_n^\pm\equiv {1\over2} (A_n\pm C^{(+)}_n)
$$
generate two {\em independent\/} $c=1/2$ Virasoro algebras, 
$$
\left[L_n^a,\,L_m^b\right] =\delta^{ab} \left(n-m\right)\,L^b_{n+m}
+{1\over24}n\,(n^2-1)\,\delta_{n+m,0}\,\delta^{ab}, 
$$
and the Hamiltonian~(\ref{eq:csg-hamilt-two}) becomes 
$$
{\cal H}_{\chi SG}^{\bh^2=4}=({2\pi/ L})\,\left({v_+}L^+_0
  + v_- L^-_0\right),\quad v_\pm=v_{\rm n}\pm\lambda/\pi.
$$
Now the spectrum can be obtained directly in terms of irreducible
representations of the two $c=1/2$ Virasoro algebras.  
To this end it is convenient to write each set of operators
$L_n^a$, $a=\pm$, in terms of an independent Majorana
fermion.  In particular,
$$
{2\pi\over L}\,L_0^a={1\over
  2i}\ix :\!\chi_a\partial_x\chi_a\!:, 
$$
which takes us back to the Hamiltonian~(\ref{eq:HXSG8pi,Maj}) obtained
by a more conventional fermionization procedure.


\begin{thebibliography}{1}
  
\bibitem{renn} S.~R. Renn, Phys.\ Rev.\ Lett. 68 (1992) 658.
  
\bibitem{kfp} C.~L. Kane, M.~P.~A. Fisher, and J. Polchinski, Phys.\ 
  Rev.\ Lett.\ 72 (1994) 4129.
  
\bibitem{haldane} F.~D.~M. Haldane., Phys.\ Rev.\ Lett.\ 74 (1995) 2090.
  
\bibitem{moore&wen} J.~E. Moore and X.-G. Wen, Phys.\ Rev.\ B 57
  (1998) 10138.

\bibitem{Dijkgraaf} R. Dijkgraaf, Nucl.\ Phys.\ B 493 (1997) 588.

\bibitem{Cardy} J. Cardy, Nucl.\ Phys.\ B 389 (1993) 577.

\bibitem{raj&jackiw} R. Jackiw and R. Rajaraman, Phys.\ Rev.\ Lett.\
  54 (1985) 1219.
  
\bibitem{Ho&Coleman} A.~F. Ho and P. Coleman, cond-mat/9812441;
  cond-mat/9903197.
  
\bibitem{nps-wip} J.~D. Naud, L.~P. Pryadko, and S.~L. Sondhi,
  unpublished.
  
\bibitem{halperin} B.~I. Halperin, Helv.\ Phys.\ Acta 56 (1983) 75.
  
\bibitem{gm1} S.~M. Girvin and A.~H. Mac{D}onald, {\em in} Novel
    Quantum Liquids in Low-Dimensional Semiconductor Structures,
  ed. S. DasSarma and A. Pinczuk (Wiley, New York, 1995).
  
\bibitem{wen} X.-G. Wen, Phys.\ Rev.\ B 43 (1991) 11025;
  Phys.\ Rev.\ Lett.\ 64 (1990) 2206; 
  Int.\ J.\ Mod.\ Phys.\ B 6 (1992) 1711; 
  Adv.\ in Phys.\ 44 (1995) 405.
  
\bibitem{wen&zee} X.-G. Wen and A. Zee, Phys.\ Rev.\ B 46 (1992) 2290.
  
\bibitem{milo&read} M. Milovanovi\'{c} and N. Read, Phys.\ Rev.\ B
  53 (1996) 13559.
  
\bibitem{Sandler} N.~P. Sandler, C. de C. Chamon, E. Fradkin, Phys.\ 
  Rev.\ B 57 (1998) 12324.
  
\bibitem{kane&fisher} C.~L. Kane and M.~P.~A. Fisher, Phys.\ Rev.\ B
  51 (1995) 13449.
  
\bibitem{stone} M. Stone, Ann.\ of Phys.\ 207 (1991) 38.
  
\bibitem{vondelft} J. von Delft and H. Schoeller, Annalen Phys.\ 7 (1998) 225.
  
\bibitem{jackiw} R. Floreanini and R. Jackiw, Phys.\ Rev.\ Lett. 59 (1987) 1873.
  
\bibitem{coleman} S. Coleman, Phys.\ Rev.\ D 11 (1975) 2088.
  
\bibitem{kogut} J. Kogut, Rev.\ of Mod.\ Phys.\ 51 (1979) 659.
  
\bibitem{wiegmann} P.~B. Wiegmann, J.\ Phys.\ C 11 (1978) 1583.
  
\bibitem{ohta} T. Ohta, Prog.\ of Theor.\ Phys.\ 60 (1978) 968.
  
\bibitem{ichinose} I. Ichinose and H. Mukaida, Int.\ J.\ Mod.\ Phys.\ 
  A 9 (1994) 1043.
  
\bibitem{BYB} P. Di Francesco, P. Mathieu, and D. S\'{e}n\'{e}chal,
  Conformal Field Theory (Springer, New York, 1997).
  
\bibitem{GKO} P. Goddard, A. Kent, and D. Olive, Comm.\ Math.\ Phys.\ 
  103 (1986) 105.
  
\bibitem{Ginsparg} P. Ginsparg, {\em in} Fields, Strings, and Critical
    Phenomena, Les Houches Session XLIX, ed. E.~Br\'{e}zin and
  J.~Zinn-Justin (North-Holland, Amsterdam, 1989).
  
\bibitem{nomura} K. Nomura and D. Yoshioka, cond-mat/9904192.
  
\bibitem{Larkin} A.~M. Finkel'stein, and A.~I. Larkin, Phys.\ Rev.\ B
  47 (1993) 10461.

\bibitem{jain} J.~K. Jain, Phys.\ Rev.\ B 40 (1989) 8079.
  
\bibitem{Bateman} Bateman Manuscript Project, Higher
    Transcendental Functions, Vol. II (Krieger, Malabar, 1981).
  
\bibitem{matveev} K.~A. Matveev, Phys.\ Rev.\ B 51 (1994) 1743.
  
\bibitem{Rajaraman} R. Rajaraman, Solitons and Instantons
  (North-Holland, Amsterdam, 1982).
  
\bibitem{FDMHunpub} F.~D.~M. Haldane, unpublished.

\bibitem{FDMH80} F.~D.~M. Haldane, Phys.\ Rev.\ Lett.\ 45 (1980) 1358.

\bibitem{black&emery} J.~L. Black and V.~J. Emery, Phys.\ Rev.\ B 23 (1981) 429.

\end{thebibliography}
\end{document}